\documentclass[12pt,pre,aps,showpacs,preprint]{revtex4}

\usepackage{graphics}
\usepackage{graphicx}
\usepackage{amsfonts}

\begin{document}

\title{Modeling Heterogeneous Materials via Two-Point Correlation Functions: I. Basic Principles}

\author{Y. Jiao}


\affiliation{\emph{Department of Mechanical and Aerospace Engineering}, \emph{Princeton University},
Princeton NJ 08544}

\author{F. H. Stillinger}


\affiliation{\emph{Department of Chemistry}, \emph{Princeton University},
Princeton NJ 08544}

\author{S. Torquato}

\email{torquato@electron.princeton.edu}

\affiliation{\emph{Department of Chemistry}, \emph{Princeton University},
Princeton NJ 08544}

\affiliation{\emph{Princeton Institute for the Science and Technology of
Materials, Princeton University}, Princeton NJ 08544}

\affiliation{\emph{Program in Applied and Computational Mathematics},
\emph{Princeton
University}, Princeton NJ 08544}

\affiliation{\emph{Princeton Center for Theoretical Physics, Princeton
University}, Princeton NJ 08544}

\date{\today}

\begin{abstract}

Heterogeneous materials abound in nature and man-made situations. Examples include 
porous media, biological materials, and composite materials. Diverse and interesting 
properties exhibited by these materials result from their complex microstructures, 
which also make it difficult to model the materials. Yeong and Torquato 
[Phys. Rev. E {\bf 57}, 495 (1998)] introduced a stochastic optimization technique 
that enables one to generate realizations of heterogeneous materials from a 
prescribed set of correlation functions. In this first part of a series of two 
papers, we collect the known necessary conditions on the standard two-point 
correlation function $S_2({\bf r})$ and formulate a new conjecture. In particular, 
we argue that given a complete two-point correlation function space, $S_2({\bf r})$ 
of any statistically homogeneous material can be expressed through a map on a 
selected set of bases of the function space. We provide new examples of realizable 
two-point correlation functions and suggest a set of analytical basis functions. 
We also discuss an exact mathematical formulation of the (re)construction problem 
and prove that $S_2({\bf r})$ cannot completely specify a two-phase heterogeneous 
material alone. Moreover, we devise an efficient and isotropy-preserving construction
 algorithm, namely, the Lattice-Point algorithm to generate realizations of materials
 from their two-point correlation functions based on the Yeong-Torquato technique.
 Subsequent analysis can be performed on the generated images to obtain desired 
macroscopic properties. These developments are integrated here into a general scheme 
that enables one to model and categorize heterogeneous materials via two-point 
correlation functions. We will mainly focus on the basic principles in this paper. 
The algorithmic details and applications of the general scheme are given in the 
second part of this series of two papers.  

\end{abstract}


\maketitle

\section{Introduction}

A heterogeneous material (medium) is one that composed of domains of different 
materials or phases (e.g., a composite) or the same material in different states 
(e.g., a polycrystal). Such materials are ubiquitous; examples include sandstones, 
granular media, animal and plant tissue, gels, foams and concrete. The microstructures
 of heterogeneous materials can be only characterized statistically via various 
types of $n$-point correlation functions \cite{1torquato}. The effective transport, 
mechanical, and electromagnetic properties of heterogeneous materials are known to 
be dependent on an infinite set of correlation functions that statistically 
characterize the microstructure \cite{1torquato}. 

\textit{Reconstruction} of heterogeneous materials from a knowledge of limited 
microstructural information (a set of lower-order correlation functions) is an 
intriguing inverse problem \cite{2PhysRevE.57.495, 3PhysRevE.58.224, 4cule:3428, 
5sheehan:53}. An effective reconstruction procedure enables one to generate accurate 
structures and subsequent analysis can be performed on the image to obtain 
macroscopic properties of the materials; see, e.g., Ref. \cite{Cu06}.
This provides a nondestructive means of 
estimating the macroscopic properties: a problem of important technological 
relevance. Another useful application is reconstruction of a three-dimensional 
structure of the heterogeneous material using information extracted from 
two-dimensional plane cuts through the material \cite{3PhysRevE.58.224}. Such 
reconstructions are of great value in a wide variety of fields, including petroleum 
engineering, biology and medicine, because in many cases one only has two-dimensional 
information such as a micrograph or image. Generating realizations of heterogeneous 
materials from a set of hypothetical correlation functions is often referred to as 
a \textit{construction} problem. A successful means of construction enables one to 
identify and categorize materials based on their correlation functions. One can also 
determine how much information is contained in the correlation functions and test 
realizability of various types of hypothetical correlation functions. Furthermore, 
an effective (re)construction procedure can be employed to investigate any physical 
phenomena where the understanding of spatiotemporal patterns is fundamental, such as in 
turbulence \cite{1torquato,16turbulence}.

A popular (re)construction procedure is  based on the use of Gaussian random 
fields: successively passing a normalized uncorrelated random Gaussian field through 
a linear and then a nonlinear filter to yield the discrete values representing the 
phases of the structure. The mathematical background used in the statistical 
topography of Gaussian random fields was originally established in the work of 
Rice \cite{17Rice, 18RandField}. Many variations of this method have been developed 
and applied since then \cite{19Quiblier, 19Berk, 19Teubner, 19roberts}. The 
Gaussian-field approach assumes that the spatial statistics of a two-phase random medium can 
be completely described by specifying only the volume fraction and standard two-point 
correlation function $S_2({\bf r})$, which gives the probability of finding two 
points separated by vector distance ${\bf r}$ in one of the phases \cite{1torquato}. 
However, to reproduce Gaussian statistics it is not enough to impose conditions on 
the first two cumulants only, but also to simultaneously ensure that higher-order 
cumulants vanish \cite{20vanKampen}. In addition, the method is not suitable for 
extension to non-Gaussian statistics, and hence is model dependent.

Recently, Torquato and coworkers have introduced another stochastic (re)construction 
technique \cite{21Rintoul, 2PhysRevE.57.495, 3PhysRevE.58.224, 4cule:3428, 
5sheehan:53}. In this method, one starts with a given, arbitrarily chosen, initial 
configuration of random medium and a set of target functions. The medium can be a 
dispersion of particle-like building blocks \cite{21Rintoul} or, more generally, 
a digitized image \cite{2PhysRevE.57.495, 3PhysRevE.58.224, 4cule:3428, 5sheehan:53}. 
The target functions describe the desirable statistical properties of the medium of 
interest, which can be various correlation functions taken either from experiments or 
theoretical considerations. The method proceeds to find a realization (configuration) 
in which calculated correlation functions best match the target functions. This is 
achieved by minimizing the sum of squared differences between the calculated and 
target functions via stochastic optimization techniques, such as the simulated 
annealing method \cite{science}. This method is applicable to multidimensional and 
multiphase media, and is highly flexible to include any type and number of 
correlation functions as microstructural information. It is both a generalization 
and simplification of the aforementioned Gaussian-field (re)construction technique. 
Moreover, it does not depend on any particular statistics \cite{1torquato}.

There are many different types of statistical descriptors that can be chosen as 
target functions \cite{1torquato}; the most basic one is the aforementioned two-point 
correlation function $S_2({\bf r})$, which is obtainable from small-angle X-ray 
scattering \cite{11debye:518}. However, not every hypothetical two-point correlation 
function corresponds to a realizable two-phase medium \cite{1torquato}. Therefore, it 
is of great fundamental and practical importance to determine the necessary 
conditions that realizable two-point correlation functions must possess 
\cite{6torquato:00, 7torquato:06}. Shepp showed that convex combinations and products 
of two scaled autocovariance functions of one-dimensional media (equivalent to 
two-point correlation functions; see definition below) satisfy all known necessary 
conditions for a realizable scaled autocovariance function \cite{8shepp}. More 
generally, we will see that a hypothetical function obtained by a particular 
combination of a set of realizable scaled autocovariance functions corresponding 
to $d$-dimensional media is also realizable.

In this paper, we generalize Shepp's work and argue that given a complete two-point 
correlation function space, $S_2({\bf r})$ of any statistically homogeneous material 
can be expressed through a map on a selected set of bases of the function space. We 
collect all known necessary conditions of realizable two-point correlation functions 
and formulate a new conjecture. We also provide new examples of realizable two-point 
correlation functions and suggest a set of analytical basis functions. We further 
discuss an exact mathematical formulation of the (re)construction problem and show 
that $S_2({\bf r})$ cannot completely specify a two-phase heterogeneous material 
alone, apart from the issue of chirality. Moreover, we devise an efficient and 
isotropy-preserving construction algorithm to generate realizations of materials from 
their two-point correlation functions. Subsequent analysis can be performed on the 
generated images to estimate desired macroscopic properties that depend on 
$S_2({\bf r})$, including both linear \cite{1torquato, A, B, C, D, E, F} and 
nonlinear \cite{G, H} behavior. These developments are integrated here into a general 
scheme that enables one to model and categorize heterogeneous materials via two-point 
correlation functions. Although the general scheme is applicable in any
space dimension, we will mainly focus on two-dimensional media here. 
In the second part of this series of two papers \cite{YJIAO},
we will provide algorithmic details and applications of our general scheme.

The rest of this paper is organized as follows: In Sec. II, we briefly introduce the 
basic quantities used in the description of two-phase random media. In Sec. III, we 
gather all the known necessary conditions for realizable two-point correlation 
functions and make a conjecture on a new possible necessary condition based on 
simulation results. In Sec. IV, we propose a general form through which the scaled 
autocovariance functions can be expressed by a set of chosen basis functions and 
discuss the choice of basis functions. In Sec. V, we formulate the (re)construction 
problem using rigorous mathematics and show that $S_2({\bf r})$ alone cannot 
completely specify a two-phase random medium. Thus, it is natural to solve the 
problem by stochastic optimization method (i.e., simulated annealing). The 
optimization procedure and the Lattice-Point algorithm are also discussed. In Sec. VI, 
we provide several illustrative examples. In Sec. VII, we make concluding remarks.   

\section{Definitions of $n$-Point Correlation Functions}

The ensuing discussion leading to the definitions of the $n$-point correlation 
functions follows closely the one given by Torquato \cite{1torquato}. Consider a 
realization of a two-phase random heterogeneous material within $d$-dimensional 
Euclidean space $\mathbb{R}^d$. To characterize this binary system, in which each phase has 
volume fraction $\phi_i$ ($i=1,~2$), it is customary to introduce the indicator 
function $I^{(i)}({\bf x})$ defined as

\begin{equation}
\label{eq101}
I^{(i)}({\bf x}) = \left\{
{\begin{array}{*{20}c}
{1, \quad\quad {\bf x} \in V_i,}\\
{0, \quad\quad {\bf x} \in \bar{V_i},}
\end{array} }\right.
\end{equation}

\noindent where $V_i\in \mathbb{R}^d$ is the region occupied by phase $i$ and $\bar{V_i}\in \mathbb{R}^d$ is the 
region occupied by the other phase. The statistical characterization of the spatial 
variations of the binary system involves the calculation of $n$-point correlation 
functions:

\begin{equation}
\label{eq102}
S^{(i)}_n({\bf x}_1,{\bf x}_2,\cdots,{\bf x}_n) = \left\langle{I^{(i)}({\bf x}_1)I^{(i)}({\bf x}_2)\cdots I^{(i)}({\bf x}_n) }\right\rangle,
\end{equation}

\noindent where the angular brackets $\left\langle{\cdots}\right\rangle$ denote 
ensemble averaging over independent realizations of the random medium. 

For \textit{statistically homogeneous} media, the $n$-point correlation function 
depends not on the absolute positions but on their relative displacements, i.e.,

\begin{equation}
\label{eq1002}
S^{(i)}_n({\bf x}_1,{\bf x}_2,\cdots,{\bf x}_n) = S^{(i)}_n({\bf x}_{12},\cdots,{\bf x}_{1n}),
\end{equation}

\noindent for all $n \ge 1$, where ${\bf x}_{ij}={\bf x}_j-{\bf x}_i$. Thus, there 
is no preferred origin in the system, which in Eq.~(\ref{eq1002}) we have chosen to 
be the point ${\bf x}_1$. In particular, the one-point correlation function is a 
constant everywhere, namely, the volume fraction $\phi_i$ of phase $i$, i.e.,

\begin{equation}
\label{eq103}
S^{(i)}_1 = \left\langle{I^{(i)}({\bf x})}\right\rangle = \phi_i,
\end{equation}

\noindent and it is the probability that a randomly chosen point in the medium 
belongs to phase $i$. For \textit{statistically isotropic} media, the $n$-point 
correlation function is invariant under rigid-body rotation of the spatial 
coordinates. For $n \le d$, this implies that $S^{(i)}_n$ depends only on the 
distances $x_{ij}=|{\bf x}_{ij}|$ ($1 \le i < j \le n$). For $n \ge d+1$, it is 
generally necessary to retain vector variables because of chirality of the medium.

The two-point correlation function $S^{(i)}_2({\bf x}_1,{\bf x}_2)$ defined as

\begin{equation}
\label{eq105}
S^{(i)}_2({\bf x}_1,{\bf x}_2) = \left\langle{I^{(i)}({\bf x}_1)I^{(i)}({\bf x}_2)}\right\rangle,
\end{equation}

\noindent is one of the most important statistical descriptors of random media. It 
also can be interpreted as the probability that two randomly chosen points 
${\bf x}_1$ and ${\bf x}_2$ both lie in phase $i$. For \textit{statistical 
homogeneous} and \textit{isotropic} media, $S^{(i)}_2$ only depends on scalar 
distances, i.e.,

\begin{equation}
\label{eq106}
S^{(i)}_2({\bf x}_1,{\bf x}_2) = S^{(i)}_2(|{\bf r}|),
\end{equation}

\noindent where ${\bf r}={\bf x}_{12}$.

Global information about the surface of the $i$th phase may be obtained by ensemble 
averaging the gradient of $I^{(i)}({\bf x})$. Since $\nabla I^{(i)}({\bf x})$ is 
different from zero only on the interfaces of the $i$th phase, the corresponding 
specific surface $s_i$ defined as the total area of the interfaces divided by the 
volume of the medium is given by \cite{1torquato}

\begin{equation}
\label{eq104}
s_i = \left\langle{|\nabla I^{(i)}({\bf x})|}\right\rangle.
\end{equation}

\noindent Note that there are other higher-order surface correlation functions which 
are discussed in detail by Torquato \cite{1torquato}.

The calculation of higher-order correlation functions encounters both analytical and 
numerical difficulties, and very few experimental results needed for comparison 
purposes are available so far. However, their importance in the description of 
collective phenomena is indisputable. A possible pragmatic approach is to study more 
complex lower-order correlation functions; for instance, the two-point cluster 
function $C^{(i)}({\bf x}_1,{\bf x}_2)$ defined as the probability that two randomly 
chosen points ${\bf x}_1$ and ${\bf x}_2$ belong to the same cluster of phase $i$ 
\cite{Cluster}; or the lineal-path function $L^{(i)}({\bf x}_1,{\bf x}_2)$ defined as 
the probability that the entire line segment between points ${\bf x}_1$ and 
${\bf x}_2$ lies in phase $i$ \cite{Lu89}. $C^{(i)}({\bf x}_1,{\bf x}_2)$ and 
$L^{(i)}({\bf x}_1,{\bf x}_2)$ of the reconstructed media are sometimes computed to 
study the non-uniqueness issue of the reconstruction 
\cite{2PhysRevE.57.495, 3PhysRevE.58.224, 4cule:3428, 5sheehan:53}.  

\section{Necessary Conditions on the Two-Point Correlation Function}

The task of determining the necessary and sufficient conditions that 
$S^{(i)}_2({\bf r})$ must possess is very complex. In the context of stochastic 
processes in time (one-dimensional processes), it has been shown that the 
autocovariance functions must not only meet all the necessary conditions we will 
present in this section but another condition on ``corner-positive'' matrices 
\cite{mcmillan}. Since little is known about corner-positive matrices, this theorem 
is very difficult to apply in practice. Thus, when determining whether a hypothetical 
function is realizable or not, we will first check all the necessary conditions 
collected here and then use the construction technique to generate realizations of 
the random medium associated with the hypothetical function as further verification.

\subsection{Known Necessary Conditions}

Here we collect all of the known necessary conditions on $S_2$ 
\cite{1torquato, 6torquato:00, 7torquato:06, 8shepp}. For a two-phase statistically 
homogeneous medium, the two-point correlation function for phase 2 is simply related 
to the corresponding function for phase 1 via the expression

\begin{equation}
\label{eq107}
S^{(2)}_2({\bf r}) = S^{(1)}_2({\bf r}) - 2\phi_1 + 1,
\end{equation}

\noindent and the \textit{autocovariance} function

\begin{equation}
\label{eq108}
\chi({\bf r}) \equiv  S^{(1)}_2({\bf r}) - {\phi_1}^2 =  S^{(2)}_2({\bf r}) - {\phi_2}^2,
\end{equation}

\noindent for phase 1 is equal to that for phase 2. Generally, for ${\bf r}=0$,

\begin{equation}
\label{eq109}
 S^{(i)}_2({\bf 0}) = \phi_i ,
\end{equation}

\noindent and in the absence of any \textit{long-range} order,

\begin{equation}
\label{eq110}
\lim_{|{\bf r}|\rightarrow \infty}  S^{(i)}_2({\bf r}) \rightarrow {\phi_i}^2 .
\end{equation}

An important necessary condition of realizable $S^{(i)}_2({\bf r})$ for a two-phase 
statistically homogeneous medium with $d$ dimensions is that the $d$-dimensional 
Fourier transform of the autocovariance function $\chi ({\bf r})$, denoted by 
$\widetilde{\chi} ({\bf k})$ must be non-negative for all wave vectors 
\cite{1torquato}, i.e., for all ${\bf k}$ 

\begin{equation}
\label{eq111}
\widetilde{\chi} ({\bf k}) = \int_{\mathbb{R}^d} \chi ({\bf r}) e^{-i{\bf k \cdot r}} {\rm d} {\bf r} \ge 0. 
\end{equation}

\noindent This non-negativity result is sometimes called the Wiener-Khintchine 
condition, which physically results since $\widetilde{\chi} ({\bf k})$ is proportional
 to the scattered radiation intensity. The two-point correlation function must satisfy
 the following bounds for all ${\bf r}$

\begin{equation}
\label{eq112}
0 \le S^{(i)}_2({\bf r}) \le \phi_i, 
\end{equation}

\noindent and the corresponding bounds on the autocovariance function are given by
\begin{equation}
\label{eq113}
-\min ({\phi_1}^2,{\phi_2}^2) \le \chi ({\bf r}) \le \phi_1\phi_2. 
\end{equation}

\noindent A corollary of Eq.~(\ref{eq113}) recently derived by Torquato 
\cite{7torquato:06} states that the infimum of any two-point correlation function of 
a statistically homogeneous medium must satisfy the inequalities

\begin{equation}
\label{eq1132}
\max\left({0, 2\phi_i -1}\right) \le \inf \left[{S^{(i)}_2({\bf r})}\right] \le \phi_i^2.
\end{equation}

Another necessary condition on $S^{(i)}_2({\bf r})$ in the case of statistically 
homogeneous and isotropic media, i.e., when $S^{(i)}_2({\bf r})$ is dependent only 
the distance $r \equiv |{\bf r}|$, is that its derivative at $r = 0$ is strictly 
negative for all $0<\phi_i<1$:

\begin{equation}
\label{eq114}
\frac{{\rm d}S^{(i)}_2}{{\rm d}r} |_{r=0} = \frac{{\rm d}\chi}{{\rm d}r} |_{r=0} < 0.
\end{equation}

\noindent This is a consequence of the fact that slope at $r=0$ is proportional to 
the negative of the specific surface \cite{1torquato}. Taking that it is axiomatic 
that $S^{(i)}_2(|{\bf r}|)$ is an even function, i.e., 
$S_2^{(i)}(|{\bf r}|)=S_2^{(i)}(-|{\bf r}|)$, then it is non-analytic at the origin.

A lesser-known necessary condition for statistically homogeneous media is the 
so-called ``triangular inequality'' that was first derived by Shepp \cite{8shepp} and 
later rediscovered by Matheron \cite{9matheron}:

\begin{equation}
\label{eq115}
 S^{(i)}_2({\bf r}) \ge  S^{(i)}_2({\bf s}) +  S^{(i)}_2({\bf t}) - \phi_i ,
\end{equation}

\noindent where ${\bf r} = {\bf t}- {\bf s}$. Note that if the autocovariance 
$\chi({\bf r})$ of a statistically homogeneous and isotropic medium is monotonically 
decreasing, nonnegative and convex (i.e., ${\rm d}^2 \chi/ {\rm d}^2 r \ge 0$), then 
it satisfies the triangular inequality Eq.~(\ref{eq115}). The triangular inequality 
implies several point-wise conditions on the two-point correlation function. For 
example, for statistically homogeneous and isotropic media, the triangular inequality 
implies the condition given by Eq.~(\ref{eq114}), the fact that the steepest descent 
of the two-point correlation function occurs at the origin \cite{8shepp}:

\begin{equation}
\label{eq116}
\left | {\frac{{\rm d}S^{(i)}_2(r)}{{\rm d}r}|_{r=0}}\right | \ge \left |{\frac{{\rm d}S^{(i)}_2(r)}{{\rm d}r}}\right |, 
\end{equation}

\noindent and the fact that $S^{(i)}_2(r)$ must be convex at the origin \cite{10Markov}:

\begin{equation}
\label{eq117}
\frac{{\rm d}^2S^{(i)}_2}{{\rm d}r^2} |_{r=0} = \frac{{\rm d}^2\chi}{{\rm d}r^2} |_{r=0} \ge 0.
\end{equation} 

Torquato \cite{7torquato:06} showed that the triangular inequality is actually a 
special case of the more general condition :

\begin{equation}
\label{eq118}
\sum\limits_{i=1}^m\sum\limits_{j=1}^m \varepsilon_i\varepsilon_j\chi({\bf r}_i-{\bf r}_j)\ge 1,
\end{equation}

\noindent where $\varepsilon_i = \pm 1$ ($i = 1,...,m$ and $m$ is odd). Note that by 
choosing $m = 3$; $\varepsilon_1\varepsilon_2 = 1$, $\varepsilon_1\varepsilon_3 = 
\varepsilon_2\varepsilon_3 = -1$, Eq.~(\ref{eq115}) can be rediscovered. If $m = 3$; 
$\varepsilon_1\varepsilon_2 = \varepsilon_1\varepsilon_3 = \varepsilon_2\varepsilon_3 
= 1$ are chosen instead, another ``triangular inequality'' can be obtained, i.e.,

\begin{equation}
\label{eq1151}
 S^{(i)}_2({\bf r}) \ge  -S^{(i)}_2({\bf s})- S^{(i)}_2({\bf t})+(4\phi_i^2 - \phi_i),
\end{equation}

\noindent where ${\bf r} = {\bf t}- {\bf s}$. Equation~(\ref{eq1151}) was first 
derived by Quintanilla \cite{quintanilla}.

 Equation~(\ref{eq118}) is a much stronger necessary condition that implies that there
 are other necessary conditions beyond those identified thus far. However, 
Eq.~(\ref{eq118}) is difficult to check in practice, because it does not have a simple
 spectral analog. One possible method is to randomly generate a set of $m$ points and 
compute the value of $\chi_{ij} = \chi ({\bf r}_i-{\bf r}_j)$. Among these values of 
$\chi_{ij}$, select the largest $m$ ones and set their coefficients 
$\varepsilon_i\varepsilon_j$ equal to $-1$. Thus, we have $m$ equations for 
$m$ $\varepsilon_i$'s. Then we can substitute the solved $\varepsilon_i$'s into 
Eq.~(\ref{eq118}) and check the inequality. If the inequality holds, then we can 
generate several different sets of random points and test the inequality in the same 
way.

\begin{figure}[bthp]
\begin{center}
$\begin{array}{c}\\
\includegraphics[height=6cm,keepaspectratio]{hypo_S2.eps}\\
\mbox{\bf (a)} \\\\\\
\includegraphics[width=5cm,keepaspectratio]{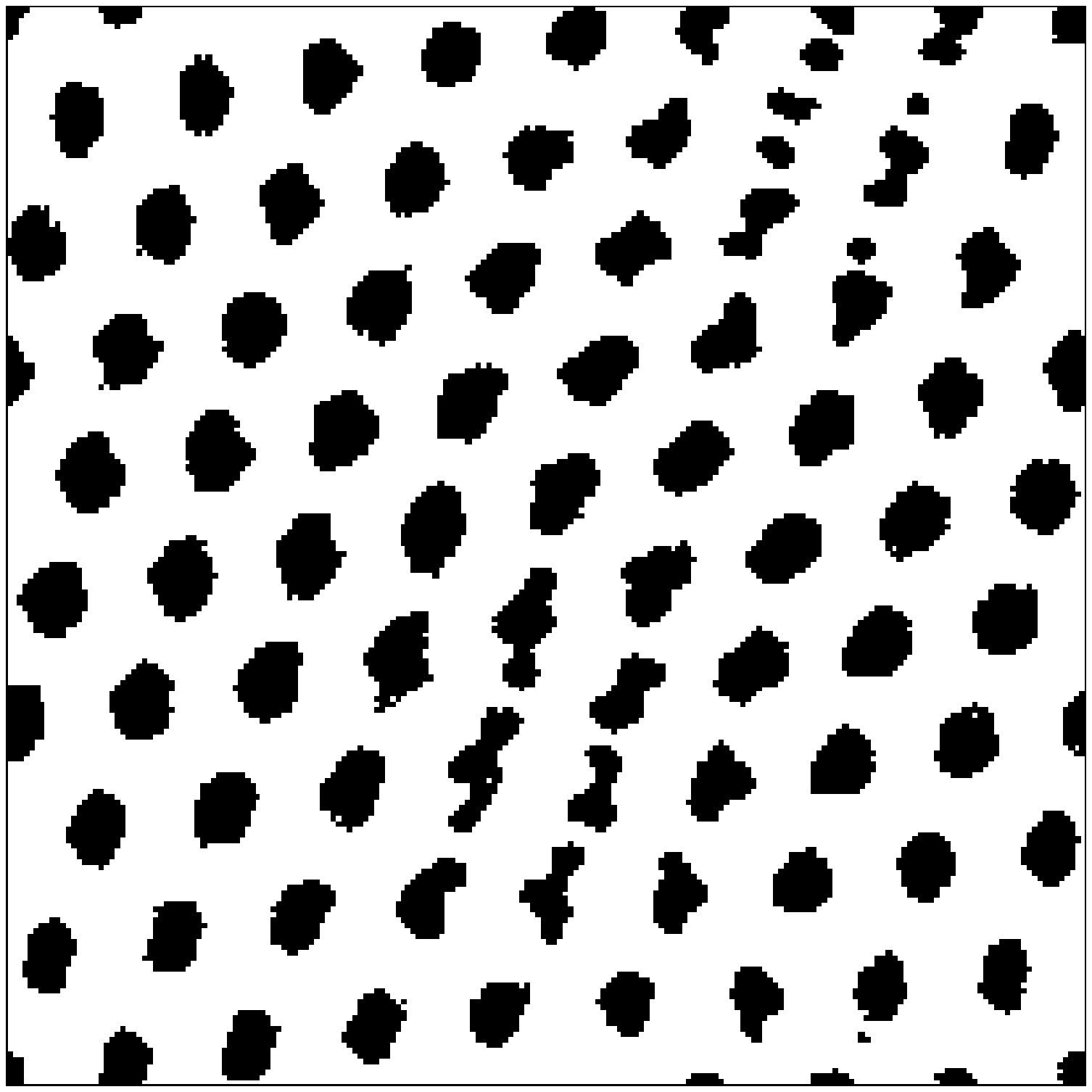}\\
\mbox{\bf (b)}
\end{array}$
\end{center}
\caption{Numerical support of the conjecture: (a) Two-point correlation functions of 
black phase for hypothetical and constructed medium. (b) Constructed medium for which 
$S_2^B$ best matches the target one. The linear size of the system $N = 200$ (pixels), volume fraction 
of black pixels $\phi_1 = 0.227$.}
\label{fig1}
\end{figure}

\subsection{Conjecture On A New Necessary Condition}

Besides the aforementioned explicit necessary conditions, we find in two-dimensional 
simulations that the value of the second peak of a nonmonotonic $S_2(r)$ for a 
statistically homogeneous and isotropic two-dimensional medium is always smaller than 
that of the medium composed of circular disks on a triangular lattice at fixed volume 
fractions, i.e.,

\begin{equation}
\label{eq119}
 S^{(i)}_2(r_p; \phi_i) \le  S^{(i)}_2(r'_p; \phi_i)^{tri},
\end{equation}

\noindent where $r_p$ and $r'_p$ denote the positions of the second peak for the two 
media, respectively and the superscript ``$tri$'' denotes the medium composed of disks
 on a triangular lattice. A hypothetical damped-oscillating two-point correlation 
function with an artificially higher second peak than that of the medium composed of 
disks on a triangular lattice at fixed volume fractions is tested by the construction 
algorithm. The results in Fig.~\ref{fig1} show that the structure for which 
$S_2^B$ (``$B$'' denotes the black phase) best matches the target function indeed has 
its ``particles'' arranged on triangular lattice  while the second peak of the target 
still cannot be reached. 

Here we make the conjecture that for any $d$-dimensional statistically homogeneous and
 isotropic medium with a two-point correlation function $S_2(r)$ that is nonmonotonic 
in $r$, the value of the first peak of its $S^{(i)}_2(r)$ away from origin is bounded 
from above by the value of the first peak of the two-point correlation function 
associated with the densest packings of $d$-dimensional identical hard spheres at 
fixed a volume fraction, i.e.,

\begin{equation}
\label{eq1192}
 S^{(i)}_2(r_p; \phi_i) \le  S^{(i)}_2(r'_p; \phi_i)^{cps},
\end{equation}

\noindent where the superscript ``$cps$'' denotes closest packings of spheres, for the
 first three dimensions, they are regular array of hard rods, hard disks on triangular
 lattice, and hard spheres on face-centered cubic lattice, respectively. For $d=4$ and
 $d=5$, the densest packings are believed to be four- and five-dimensional 
checkerboard lattice packing, respectively \cite{conway}. Note that this conjectured 
condition could be a corollary of Eq.~(\ref{eq118}) or some other unknown necessary 
conditions.
 
\section{Modeling Two-Point Correlation Function via Basis Functions}

\subsection{Combination of Realizable Two-Point Correlation Functions} 

It is first shown by Shepp \cite{8shepp} that the convex combination and product of 
two realizable \textit{scaled autocovariance functions} for \textit{one-dimensional} 
statistically homogeneous media satisfy all known necessary conditions, i.e.,

\begin{equation}
\label{eq125}
\begin{array}{l}f^c(r) = \alpha_1 f_1(r) + \alpha_2 f_2(r),\\\\
f^p(r) = f_1(r)f_2(r),
\end{array}
\end{equation} 

\noindent where $0\le \alpha_i \le 1$ ($i=1,2$), $\alpha_1+\alpha_2=1$ and the 
superscripts ``$c$'' and ``$p$'' denote ``combination'' and ``product'', respectively.
 The scaled autocovariance function $f({\bf r})$ of a statistically homogeneous 
material is defined as \cite{7torquato:06}

\begin{equation}
\label{eq120}
f({\bf r}) \equiv \frac{\chi({\bf r})}{\phi_1\phi_2} = \frac{S^{(i)}_2({\bf r})-\phi_i^2}{\phi_1\phi_2}.
\end{equation}

The necessary conditions for realizable scaled autocovariance function $f({\bf r})$ 
can be obtained from Eq.~(\ref{eq120}) and the equations by which the necessary 
conditions for realizable two-point correlation function $S^{(i)}_2({\bf r})$ are 
given. From Eqs.~(\ref{eq115}) and (\ref{eq1151}), we can obtain the triangular 
inequalities for $f({\bf r})$, respectively,

\begin{equation}
\label{eq121}
f({\bf r}) \ge f({\bf s}) + f({\bf t}) - 1.
\end{equation}

\begin{equation}
\label{eq1211}
f({\bf r}) \ge -f({\bf s}) - f({\bf t}) - 1.
\end{equation}

\noindent Moreover, the bounds of $f({\bf r})$ become

\begin{equation}
\label{eq122}
-\min\left [{\frac{\phi_1}{\phi_2},\frac{\phi_2}{\phi_1}}\right ] \le f({\bf r}) \le 1,
\end{equation}

\noindent and the corollary Eq.~(\ref{eq1132}) is equivalent to

\begin{equation}
\label{eq1222}
-\min\left[{\frac{\phi_1}{\phi_2}, \frac{\phi_2}{\phi_1}}\right] \le f_{\inf} \le 0,
\end{equation}

\noindent where $f_{\inf}$ is the infimum of $f({\bf r})$. Our focus in this paper 
will be hypothetical continuous functions $f(r)$ that are dependent only on the scalar
 distance $r = |{\bf r}|$ which could potentially correspond to statistically 
homogeneous and isotropic media without long range order, i.e., 

\begin{equation}
\label{eq123}
f(0) = 1, \quad\quad \lim_{r \rightarrow \infty}f(r) \rightarrow 0.
\end{equation}

\noindent $f(r)$ is also absolutely integrable so that the Fourier transform of 
$f(r)$ exists and is given by

\begin{equation}
\label{eq124}
\widetilde{f}(k) = (2 \pi)^{d/2} \int\limits_0^{\infty} r^{d-1}f(r)\frac{J_{(d/2)-1}(kr)}{(kr)^{(d/2)-1}}{\rm d}r \ge 0,
\end{equation}

\noindent where $k = |{\bf k}|$ and $J_v(x)$ is the Bessel function of order $v$.

 Generalization of Eq.~(\ref{eq125}) to higher dimensions is straightforward. Suppose 
$f_i(r)$ ($i=1,...,m$) are the scaled autocovariance functions for $d$-dimensional 
statistically homogeneous and isotropic media, then the convex combination $f^c(r)$ 
and product $f^p(r)$ defined as 

\begin{equation}
\label{eq126}
\begin{array}{l}
f^c(r) = \sum\limits_{i=1}^m \alpha_i f_i(r), \\\\
f^p(r) = \prod\limits_{i=1}^m f_i(r),
\end{array}
\end{equation}

\noindent satisfy all known necessary conditions, where $0 \le \alpha_i \le 1$ 
($i=1,...,m$) and $\sum_{i=1}^m \alpha_i = 1$. Equation~(\ref{eq126}) is of great 
fundamental and practical importance. On the one hand, it enables us to construct new 
realizable two-point correlation functions with properties of interest, corresponding 
to structures of interest, from a set of known functions. Thus, one can categorize 
microstructures with the set of known functions and the proper combinations. On the 
other hand, suppose that we can find a ``full'' set of those \textit{basis} scaled 
autocovariance functions $\{f_i(r)\}_{i=1}^m$, then the scaled autocovariance function
 of any statistically homogeneous and isotropic medium can be expressed in term of the
 combinations of the basis functions, i.e.,

\begin{equation}
\label{eq127}
f(r) = \wp[\{f_i(r)\}_{i=1}^m] \equiv \wp [f_1(r),f_2(r),...,f_m(r)],
\end{equation}

\noindent where $\wp$ denotes a map composed of convex combinations and products of 
$\{f_i(r)\}_{i=1}^m$. For example, for $m=5$, a \textit{possible} explicit form for 
$\wp$ is

\begin{equation}
\label{eq128}
\wp [\{f_i(r)\}_{i=1}^5] = \alpha_1 f_1(r) + \alpha_2 [\beta_1 f_2(r) + \beta_2 f_3(r)] + \alpha_3 [f_4(r)f_5(r)],
\end{equation}

\noindent where $0 \le \alpha_i, \beta_j \le 1$ and $\sum_i \alpha_i = \sum_j \beta_j
 = 1$ ($i=1,2,3; j=1,2$). Once the scaled autocovariance function (or equivalently the
 two-point correlation function) of a medium is known, an effective reconstruction 
procedure enables one to generate accurate structures at will, and subsequent analysis
 can be performed on the image to obtain desired macroscopic properties of the medium.
 In other words, the medium is actually \textit{modeled} by a set of basis scaled 
autocovariance functions $\{f_i(r)\}_{i=1}^m$ and a particular map $\wp [\{f_i(r)\}_
{i=1}^m]$. There could be different choices of the basis functions (like different 
basis choices of a Hilbert space), and we would like the basis functions to have nice 
mathematical properties, such as simple analytical forms. Let $\{f_i^0(r)\}_{i=1}^m$ 
denotes our choice of the basis functions. Thus, the media can be represented merely 
by different maps $\wp^0$'s. Note that a hypothetical two-point correlation function 
corresponds to a hypothetical map $\wp^0_h$ and effective construction algorithms can 
be used to test the realizability of $\wp^0_h$.

\subsection{Choice of Basis Functions}
A systematic way of determining the basis functions $\{f_i^0\}_{i=1}^m$ is not 
available yet. Here we take the first step to determine the bases by considering 
certain known realizable analytical two-point correlation functions and the 
corresponding scaled autocovariance functions. For convenience, we categorize these 
functions into three families: ($i$) \textit{monotonically decreasing} functions; 
($ii$) \textit{damped-oscillating} functions; and ($iii$) functions of 
\textit{known constructions}.

\begin{figure}[bthp]
\begin{center}
$\begin{array}{c}\\
\includegraphics[height=6cm,keepaspectratio]{Debye_obj2.eps}\\
\mbox{\bf (a)}\\\\\\
\includegraphics[width=5cm,keepaspectratio]{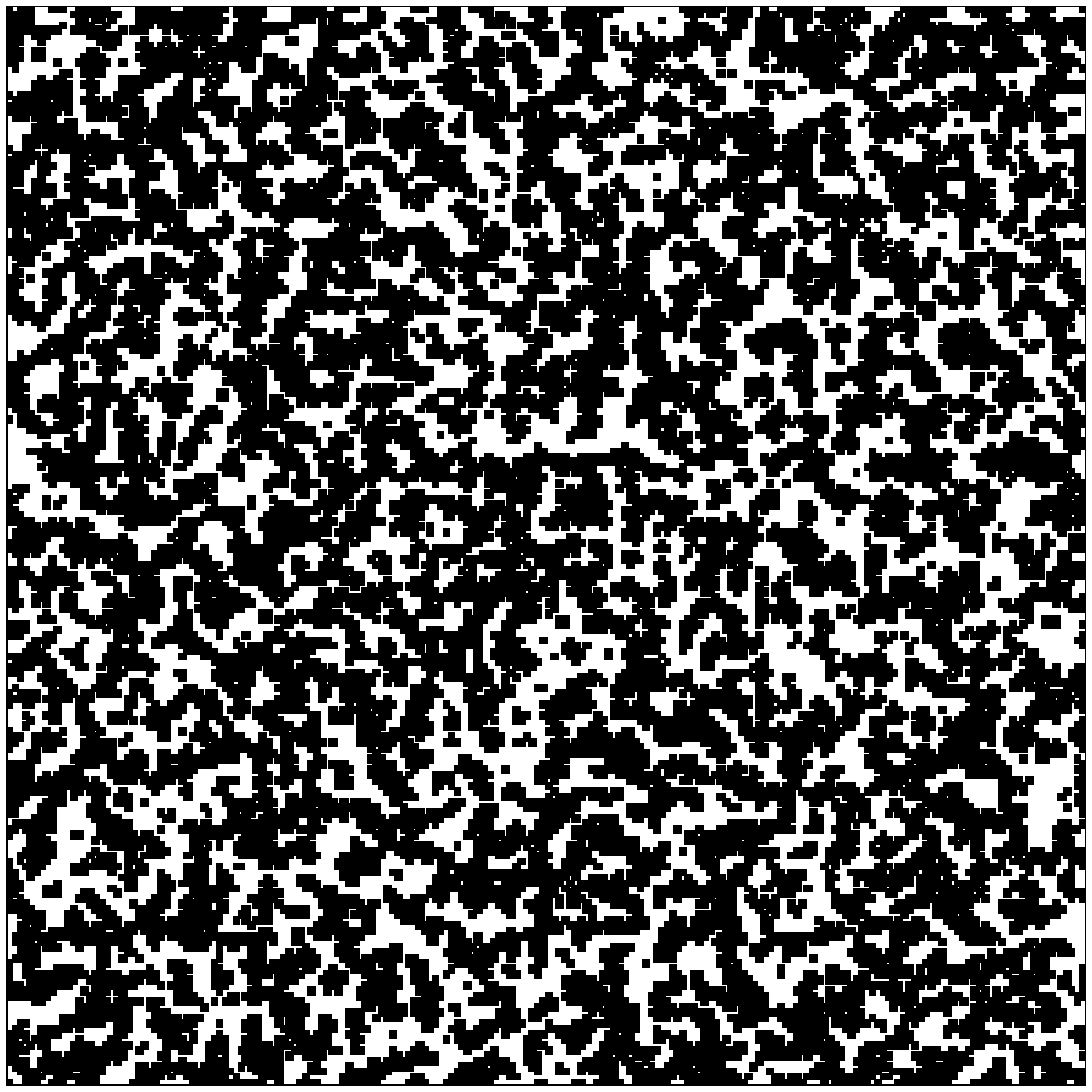}\\
\mbox{\bf (b)}
\end{array}$
\end{center}
\caption{(a) Debye random medium function $f_D(r)$ with $a = 5$. (b) A realization of 
Debye random media with the volume fractions $\phi_1 = 0.68$, $\phi_2 = 0.32$.}
\label{fig2}
\end{figure}

The family of monotonically decreasing functions includes the simple exponentially 
decreasing function introduced by Debye \cite{11debye:518} and polynomial functions. 
The former is given by

\begin{equation}
\label{eq129}
f_D(r) = \exp(-r/a),\quad r \ge 0,
\end{equation}

\noindent where $a$ is a correlation length, corresponding to structures in which one 
phase consists of ``random shapes and sizes'' \cite{11debye:518, 12debye:679} (shown 
in Fig.~\ref{fig2}). It is now known that certain types of space tessellations have 
autocovariance functions given by Eq.~(\ref{eq129}) \cite{13Stochastic}. We have 
referred to this class of structures as \textit{Debye random media} \cite{1torquato, 
2PhysRevE.57.495}. 

Another example of monotonically decreasing functions is the family of polynomials of 
order $n$ ($n \ge 1$) given by

\begin{equation}
\label{eq130}
f_P^n(r) = \left\{\begin{array}{*{20}c} (1-r/a)^n \quad 0 \le r \le a, \\\\ 0 \quad\quad r>a, \end{array}\right.
\end{equation}

\noindent where $a$ is the correlation length. The polynomial function of order 1 is 
shown to be realizable only for a statistically homogeneous two-phase medium in one 
dimension \cite{7torquato:06, 8shepp}. We have constructed for the first time 
realizations of random media in dimensions $d \le n$ that correspond to the polynomial
 function of order $n$ with very high numerical precision using the Yeong-Torquato 
construction technique \cite{2PhysRevE.57.495}. However, for dimensions higher than 
$n$, Eq.~(\ref{eq130}) violates Eq.~(\ref{eq124}). We will henceforth assume that the 
polynomial functions of order $n$ are realizable in dimensions $d \le n$.

An example of the family of damped-oscillating functions is given by 
\cite{2PhysRevE.57.495, 7torquato:06}

\begin{equation}
\label{eq131}
f_O(r) =\sum_i A_i \exp(-r/a_i)\cos(q_ir + \psi_i),\quad r \ge 0,
\end{equation}

\noindent where the parameters $A_i$ and $a_i$ ($i = 1, 2, ...$) control the amplitude
 of the $f_O$ profile, $q_i$ is the wavenumber and $\psi_i$ is the phase angle. 
In general, the wavenumber should be a function of $r$ which could correspond to the 
different distances between successive neighbor shells of a crystalline material. 
Note that $q_i$ and $\psi_i$ need to be carefully chosen such that $f_O(r)$ satisfies 
all known necessary conditions.

\begin{figure}[bthp]
\begin{center}
$\begin{array}{c}\\
\includegraphics[height=6cm,keepaspectratio]{OD_obj2.eps}\\
\mbox{\bf (a)}\\\\\\
\includegraphics[width=5cm,keepaspectratio]{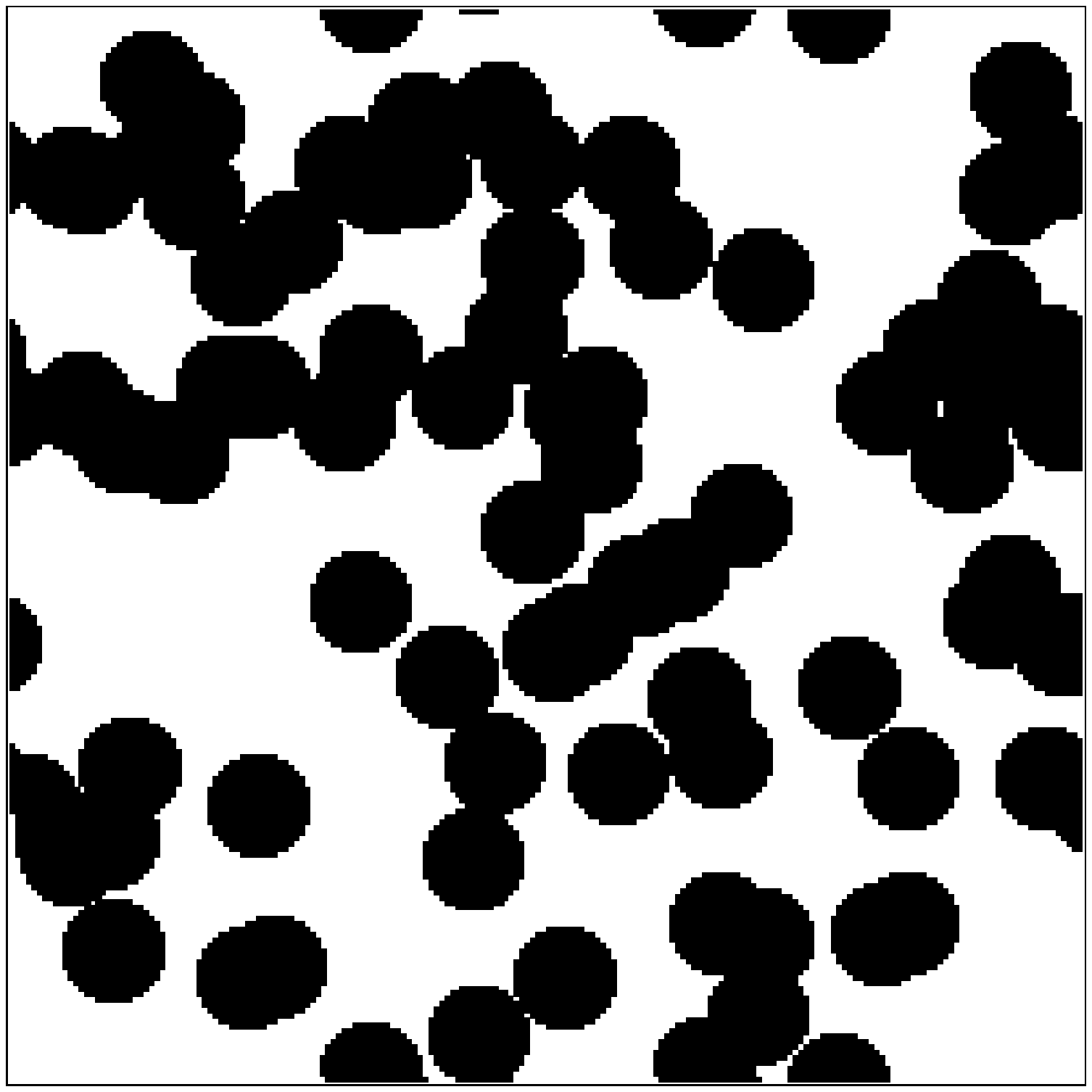}\\
\mbox{\bf (b)}
\end{array}$
\end{center}
\caption{(a) Scaled autocovariance function  $f_S(r)$ of two-dimensional identical 
overlapping disks with volume fractions $\phi_1 = 0.45,~ \phi_2 = 0.55$. (b) A 
realization of two-dimensional identical overlapping disks with volume fractions 
$\phi_1 = 0.45,~ \phi_2 = 0.55$. The radius of disks $R = 5$.}
\label{fig3}
\end{figure}

 The family of functions of known constructions includes scaled autocovariance 
functions of $d$-dimensional identical overlapping spheres \cite{1torquato, stell1983}
 and symmetric-cell materials \cite{1torquato, Lu1990}. For overlapping spheres of 
radius $R$, the scaled autocovariance function for the particle phase (spheres) is 
given by

\begin{equation}
\label{eq132}
f_S(r) = \frac{\exp[-\rho v_2(r;R)]-\phi_1^2}{\phi_1\phi_2},
\end{equation}

\noindent where $\phi_1$ and $\phi_2$ are volume fractions of the spheres and matrix 
respectively, $\rho = N/V$ is the number density of spheres, and $v_2(r; R)$ is the 
union volume of two spheres of radius $R$ whose centers are separated by $r$. For the 
first three space dimensions, the latter is respectively given by

\begin{equation}
\label{eq133}
\frac{v_2(r;R)}{v_1(R)} = 2\Theta\left ({r-2R}\right ) + \left({1+\frac{r}{2R}}\right)\Theta(2R-r), 
\end{equation}

\begin{equation}
\label{eq134}
\frac{v_2(r;R)}{v_1(R)} = 2\Theta(r-2R) + \frac{2}{\pi}\left[{\pi + \frac{r}{2R}\left({1-\frac{r^2}{4R^2}}\right)^{\frac{1}{2}} - \cos^{-1}\left({\frac{r}{2R}}\right)}\right]\Theta(2R-r),
\end{equation}

\begin{equation}
\label{eq135}
\frac{v_2(r;R)}{v_1(R)} = 2\Theta(r-2R) + \left[{1 + \frac{3r}{4R} - \frac{1}{16}\left({\frac{r}{R}}\right)^3}\right]\Theta(2R-r),
\end{equation}

\noindent where $\Theta(x)$ is the Heaviside function, and $v_1(R)$ is the volume of 
a $d$-dimensional sphere of radius $R$ given by

\begin{equation}
\label{eq136}
v_1(R) = \frac{\pi^{d/2}}{\Gamma(1+d/2)}R^d, 
\end{equation}

\noindent where $\Gamma(x)$ is the gamma function. For $d = 1, 2$ and $3$, $v_1(R) = 
2R,\pi R^2$ and $4\pi R^3/3$, respectively.

\begin{figure}[bthp]
\begin{center}
$\begin{array}{c}\\
\includegraphics[height=6cm,keepaspectratio]{CB_obj2.eps}\\
\mbox{\bf (a)}\\\\\\
\includegraphics[width=5cm,keepaspectratio]{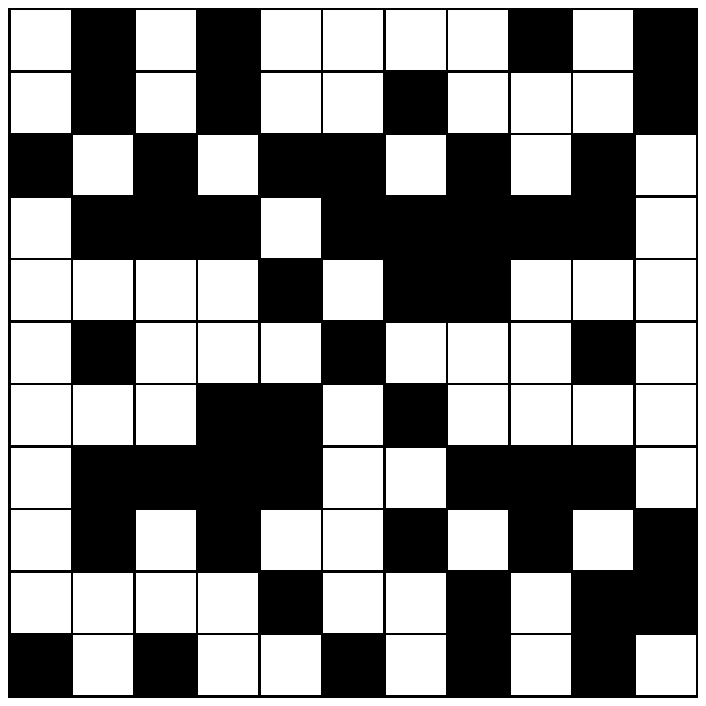}\\
\mbox{\bf (b)}
\end{array}$
\end{center}
\caption{(a) Scaled autocovariance function  $f_C(r)$ of two-dimensional random 
checkerboard. (b) A realization of two-dimensional random checkerboard. The length of 
the cells $a = 10$.}
\label{fig4}
\end{figure}

Two-phase symmetric-cell materials are constructed by partitioning space into cells of
 arbitrary shapes and sizes, with cells being randomly designated as phase 1 and 2 
with probability $\phi_1$ and $\phi_2$ from a uniform distribution \cite{1torquato}. 
For such a statistically homogeneous and isotropic medium, the scaled autocovariance 
function is given by

\begin{equation}
\label{eq137}
f_C(r) = W_2^{(1)} (r),
\end{equation}

\noindent where $W_2^{(1)}(r)$ is the probability that two points separated by 
distance $r$ are in the same cell. This quantity is only a function of cell shapes and
 sizes, depending on a dimensionless size-averaged intersection volume of two cells 
defined by

\begin{equation}
\label{eq138}
 W_2^{(1)} (r) = \frac{\left\langle{v^{int}_2(r;R)}\right\rangle_R}{\left\langle{v_1(r)}\right\rangle_R},
\end{equation}

\noindent where $R$ is the size parameter for each cell, $\left\langle{v^{int}_2(r;R)}
\right\rangle_R$ is the size-averaged intersection volume of two cells whose centers 
are separated by $r$ and $\left\langle{v_1(r)}\right\rangle_R$ is the size-averaged 
single-cell volume. The random checkerboard is a very useful model of symmetric-cell 
material because its $f_C(r)$ is known analytically for $d = 1$ and $2$ 
\cite{1torquato}. For $d = 1$, it is easy to verify that the probability of finding 
two points in the same one-dimensional cell is given by

\begin{equation}
\label{eq139}
 W_2^{(1)} (r) = \left\{{\begin{array}{*{20}c}1-r/a, \quad 0\le r\le a \\\\ 0, \quad\quad r\ge a. \end{array}}\right.
\end{equation}

\noindent where $a$ is the length of the side of a square cell. Note that 
Eq.~(\ref{eq139}) is just the polynomial function of order one [cf. Eq.~(\ref{eq130})]
 for one-dimensional homogeneous media. For $d=2$, $W_2^{(1)}(r)$ is given by

\begin{equation}
\label{eq140}
 W_2^{(1)} (r) = \left\{{\begin{array}{*{20}c}\displaystyle{ 1 + \frac{1}{\pi} \left[{\left({\frac{r}{a}}\right)^2 - 4\left({\frac{r}{a}}\right)}\right], \quad\quad 0\le r\le a,} \\\\ \displaystyle{ 1 - \frac{1}{\pi}\left[{2 + \left({\frac{r}{a}}\right)^2}\right] + \frac{4}{\pi}\left[{\sqrt{\left({\frac{r}{a}}\right)^2 - 1} - \cos^{-1}\left({\frac{a}{r}}\right)}\right], ~ a \le r \le \sqrt{2}a,}\\\\  0, \quad\quad\quad r \ge \sqrt{2} a.  \end{array}}\right.
\end{equation}

\noindent For $d = 3$, $W_2^{(1)}(r)$ is given by

\begin{equation}
\label{eq1401}
W_2^{(1)}(r) = \frac{2}{\pi}\int_0^{\pi/2}\int_0^{\pi/2}W(r, \theta, \phi)\sin\theta d\theta d\phi,
\end{equation}

\noindent where

\begin{equation}
\label{eq1402}
\begin{array}{c}
W(r, \theta, \phi) = [1-r\cos\theta][1-r\sin\theta\sin\phi][1-r\sin\theta\cos\phi]\\\quad\quad\quad\times\Theta(1-r\cos\theta)\Theta(1-r\sin\theta\sin\phi)\Theta(1-r\sin\theta\cos\phi).
\end{array}
\end{equation}

\noindent Note that $W_2^{(1)}(r)$ for $d = 3$ does not have a simple analytical form.

There are some other known scaled autocovariance functions for different types of 
materials (e.g., $d$-dimensional identical hard spheres \cite{1torquato}) and 
realizable hypothetical functions (e.g., complementary error function, see Appendix). 
However we do not include these functions in our basis function set because they do 
not have simple analytical mathematical forms. Note that some of the basis functions 
are dimension-dependent (e.g., $f_O$ and $f_C$), and thus the proper forms of basis 
functions should be used for different dimensions. In the following discussion, 
without further specification, we will focus on two-dimensional cases.

\section{Generating Realizations of Heterogeneous Materials}

Consider a digitized (i.e., pixelized) representation of a heterogeneous material. 
Different colored pixels (in a discrete coloring scheme) may have numerous 
interpretations. The image can reflect different properties, such as the geometry 
captured by a photographic image, topology of temperature and scalar velocity fields 
in fluids, distribution of magnitudes of electric and magnetic fields in the medium, 
or variations in chemo-physical properties of the medium. In the last case, typical 
examples are composite materials in which the different phases may have different 
thermal, elastic or electromagnetic properties, to name a few. 

\subsection{Exact Equations for Digitized Media}

\begin{figure}[bthp]
\begin{center}
$\begin{array}{c@{\hspace{1cm}}c}\\
\includegraphics[width=4cm,keepaspectratio]{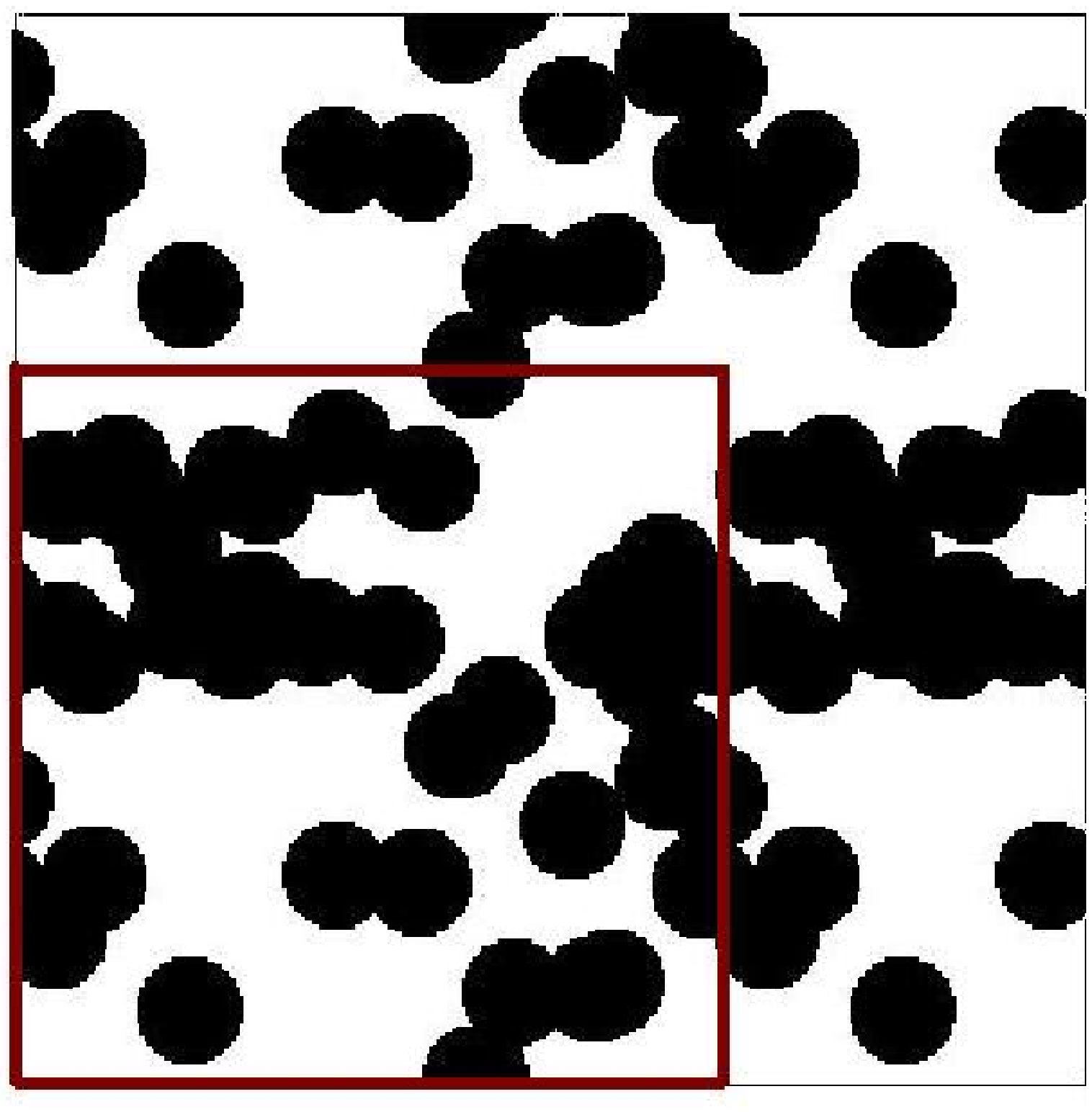} &
\includegraphics[width=4cm,keepaspectratio]{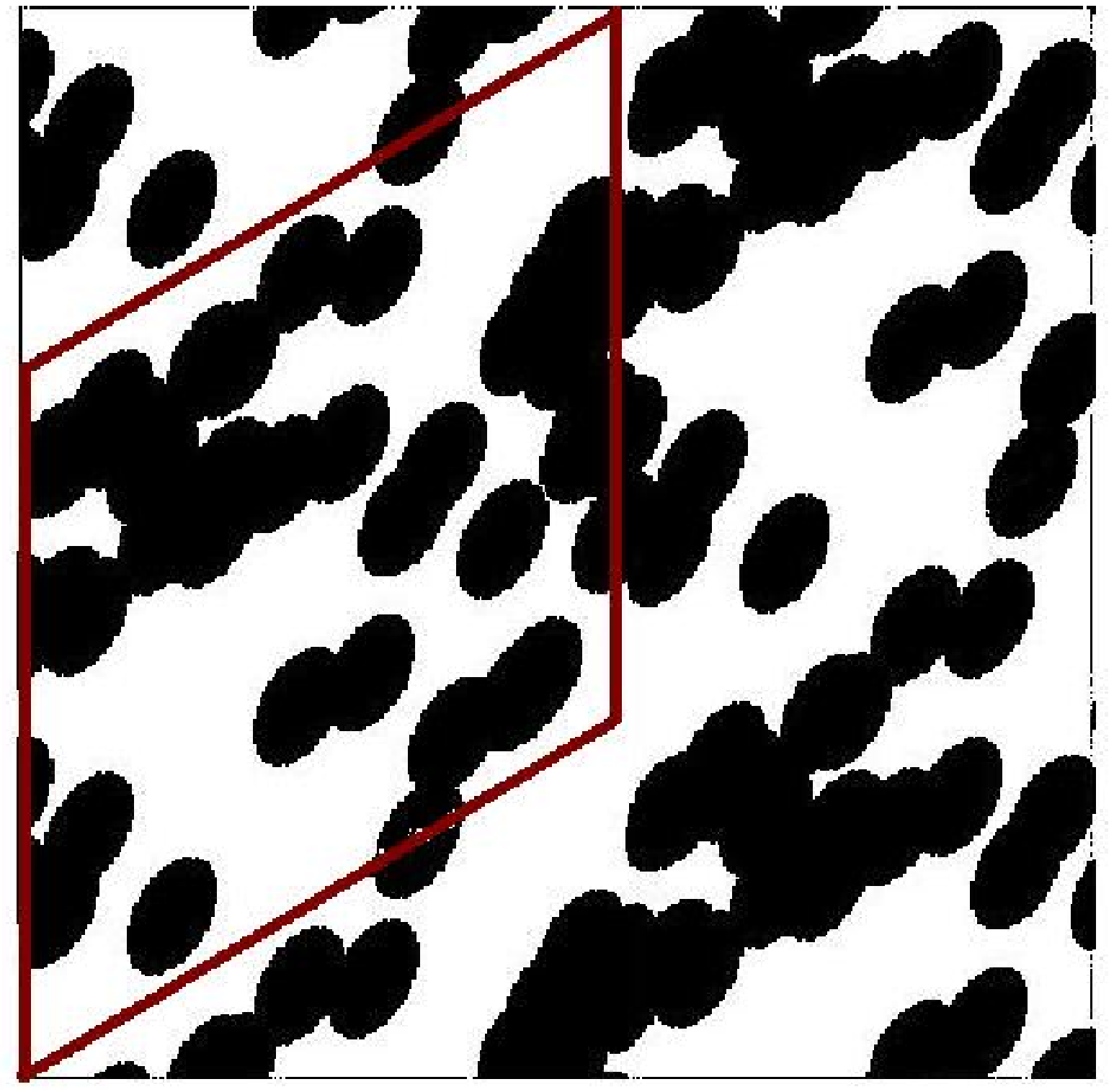} \\
\mbox{\bf (a)} & \mbox{\bf (b)}
\end{array}$
\end{center}
\caption{Different digitized media with the same arrary representation: (a) 
Overlapping disks generated on square lattice with a square unit cell. (b) Orientated 
overlapping ellipses generated on triangular lattice with a rhombical unit cell.}
\label{fig5}
\end{figure}

Our focus in this paper is two-dimensional, two-phase statistically homogeneous and 
isotropic random media composed of black and white pixels. Such a system can be 
represented as a two-dimensional array, i.e.,

\begin{equation}
\label{eq201}
{\bf I} = \left[{\begin{array}{*{20}c} I_{11} \quad I_{12} \quad ... \quad I_{1N} \\ I_{21}  \quad I_{22} \quad ... \quad I_{2N} \\ \vdots \quad\quad \ddots \quad\quad \vdots \\ I_{N1}\quad I_{N2} \quad ... \quad I_{NN} \end{array}}\right],\end{equation}

\noindent where the integer $N$ categorizes the linear size of the system ($N^2$ is 
the total number of pixels in the system) and the entries $I_{ij}$ ($i,j = 1,...,N$) 
can only take the value of $0$ or $1$, which correspond to the white and black phases,
 respectively. Note that Eq.~(\ref{eq201}) is only an abstract representation and the 
real morphological configuration of the medium also depends on the choice of 
\textit{lattices}. For example, as shown in Fig.~\ref{fig5}, the isotropic medium 
composed of overlapping disks generated on a square lattice and the anisotropic medium
 composed of orientated ellipses generated on a triangular lattice have the same 
array presentation. A vector distance in the digitized medium can be uniquely 
expressed as

\begin{equation}
\label{eq202}
{\bf r} = n_1 {\bf e}_1 + n_2 {\bf e}_2,
\end{equation}

\noindent where ${\bf e}_1$ and ${\bf e}_2$ are lattice vectors for the particular 
lattice and $n_1$, $n_2$ are integers. For example, for square lattice, ${\bf e}_1 = 
{\bf i}$, ${\bf e}_2 = {\bf j}$, where ${\bf i}$, ${\bf j}$ are unit vectors along 
horizontal and vertical directions, respectively; while for triangular lattice 
${\bf e}_1 = \sqrt{3}{\bf i} +  {\bf j}/2$, ${\bf e}_2 = {\bf j}$.

 Without loss of generality, we choose the black phase to be the phase of interest and
 assume \textit{periodic boundary condition} is applied, which is commonly used in 
computer simulations. The two-point correlation function $S_2({\bf r})$ of the black 
phase can be calculated based on its probabilistic nature, i.e., the probability of 
finding two points separated by the vector distance ${\bf r}$ in the same phase. The 
value of two-point correlation function for a particular ${\bf r} = n_1 {\bf e}_1 + 
n_2 {\bf e}_2$ is given by

\begin{equation}
\label{eq203}
S_2({\bf r}) \equiv S_2(n_1,n_2) =\frac{ \sum\limits_{i=1}^N\sum\limits_{j=1}^N I_{ij} I_{(i+n_1)(j+n_2)}}{N^2},
\end{equation}

\noindent where $I_{ij}$ are entries of ${\bf I}$ defined in Eq.~(\ref{eq201}), $n_1$ 
and $n_2$ are integers satisfying $|n_1|,|n_2| \le [N/2]$ due to \textit{minimum image
 distance} convention. For statistically isotropic media, the two-point correlation 
function only depends on the magnitude of ${\bf r}$, i.e., ${r} \equiv |{\bf r}|$, 
thus we have

\begin{equation}
\label{eq204}
S_2(r) =\displaystyle{\frac{ \sum\limits_{(m,n)\in\Omega} \left[{\sum\limits_{i=1}^N\sum\limits_{j=1}^N I_{ij} I_{(i+m)(j+n)}}\right]}{\omega N^2}},
\end{equation}

\noindent where

\begin{equation}
\label{eq205}
\Omega = \left\{{(m,n)~|~ m^2 + n^2 = {r}^2, ~ {r} \le [N/2]}\right\},
\end{equation}

\noindent and $\omega$ is the number of elements of set $\Omega$.

It is well known that the two-point correlation function cannot completely specify a 
two-phase heterogeneous material alone. Here, we provide a proof of this statement for
 two-dimensional statistically homogeneous digitized media. Note that the proof 
trivially extends to any dimension. Suppose we already know the value of 
$S_2({\bf r})$ for every vector distance ${\bf r}$, using Eq.~(\ref{eq203}), we can 
obtain a set of equations of $I_{ij}$, i.e., for each ${\bf r} = n_1 {\bf e}_1 + 
n_2 {\bf e}_2$

\begin{equation}
\label{eq206} 
 \sum\limits_{i=1}^N\sum\limits_{j=1}^N I_{ij} I_{(i+n_1)(j+n_2)} - N^2 S_2(n_1, n_2) = 0,
\end{equation}

\noindent Since the digitized medium is represented by ${\bf I}$, once we obtain all 
the entries $I_{ij}$ (the unknowns in Eq.~(\ref{eq206})), the medium is 
(re)constructed. In Eq.~(\ref{eq206}), the number of unknowns $N_u$ equals $N^2$ and 
the number of equations $N_e$ equals the number of all possible vector distances in 
the digitized medium. To calculate $N_e$, all possible different combinations of 
integers $n_1$ and $n_2$ subjected to $|n_1|,|n_2| \le [N/2]$ need to be considered. 
This quantity is given by $N_e = 4[N/2]^2 - 8[N/2] + 6$, which is smaller than the 
number of unknowns $N_u = N^2$ for normal-sized systems (i.e., $N = 10 \sim 10^3$).

Similar proof is applied to the case when we average over the angles of vector 
${\bf r}$ to yield $S_2(r)$ that depends only on the radius distance $r$. The 
angle-averaged equations of $I_{ij}$ are given by

\begin{equation}
\label{eq207} 
  \sum\limits_{(m,n)\in\Omega} \left[{\sum\limits_{i=1}^N\sum\limits_{j=1}^N I_{ij} I_{(i+m)(j+n)}}\right] - \omega N^2 S_2(r) = 0,
\end{equation}

\noindent where $\Omega$ is given by Eq.~(\ref{eq205}). The number of equations is 
even smaller for the angle-averaged case, while the number of unknowns does not 
change. The analysis shows that one could never find a unique solution of $I_{ij}$ 
from neither Eq.~(\ref{eq206}) nor Eq.~(\ref{eq207}) unless some assumptions have 
been made that reduce $N_u$ such that $N_u = N_e$. For example, in an interesting 
model to study spatial distribution of algae, the algae are put on top of each other 
in order to reduce the unknowns \cite{nature}. In general cases, one could use the 
stochastic optimization procedure (i.e., simulated annealing) \cite{science, 
2PhysRevE.57.495} to find solutions of Eq.~(\ref{eq206}) and Eq.~(\ref{eq207}). 
Note that although the aforementioned proofs focus on two-dimensional media, they 
trivially extend to any dimensions ( e.g., $d = 1, 3$).

\subsection{Stochastic Optimization Procedure}

Generally, consider a given set of correlation functions $f^{\alpha}_n({\bf r}_1,
{\bf r}_2,...,{\bf r}_n)$ of the phase of interest that provides partial information 
on the random medium. The index $\alpha$ is used to denote the type of correlation 
functions. Note that the set of $f^{\alpha}_n$ should not be confused with the basis 
function set $\wp$, the former contains correlation functions of different type, i.e.,
 two-point correlation function, lineal-path function, two-point cluster function, 
etc., while the latter contains basis functions through which the scaled 
autocovariance function of the medium of interest can be expressed. The information 
contained in $f^{\alpha}_n$ could be obtained either from experiments or it could 
represent a hypothetical medium based on simple models. In both cases we would like 
to generate the underlying micro-structure with a specified set of correlation 
functions. In the former case, the formulated inverse problem is frequently referred 
to as a ``reconstruction'' procedure, and in the latter case as a ``construction''.

As we have noted earilier, it is natural to formulate the construction or 
reconstruction problem as an optimization problem \cite{2PhysRevE.57.495, 
3PhysRevE.58.224, 4cule:3428, 5sheehan:53}. The discrepancies between the statistical 
properties of the best generated structure and the imposed ones is minimized. This 
can be readily achieved by introducing the ``energy'' function $E$ defined as a sum 
of squared differences between target correlation functions, which we denote by 
$\widehat{f}^{\alpha}_n$, and those calculated from generated structures, i.e.,

\begin{equation}
\label{eq208}
E = \sum\limits_{{\bf r}_1,{\bf r}_2,...,{\bf r}_n}\sum\limits_{\alpha}\left[{f^{\alpha}_n({\bf r}_1,{\bf r}_2,...,{\bf r}_n)-\widehat{f}^{\alpha}_n({\bf r}_1,{\bf r}_2,...,{\bf r}_n)}\right]^2.  
\end{equation}

\noindent Note that for every generated structure (configuration), these is a set of 
corresponding $f^\alpha_n$. If we consider every structure (configuration) as 
a ``state'' of the system, $E$ can be considered as a function of the states. The 
optimization technique suitable for the problem at hand is the method of simulated 
annealing \cite{science}. It is a popular method for the optimization of large-scale 
problems, especially those where a global minimum is hidden among many local extrema. 
The concept of finding the lowest energy state by simulated annealing is based on a 
well-known physical fact: If a system is heated to a high temperature $T$ and then 
slowly cooled down to absolute zero, the system equilibrates to its ground state. At 
a given temperature $T$, the probability of being in a state with energy $E$ is given 
by the Boltzmann distribution $P(E) \sim \exp(-E/T)$. At each annealing step $k$, the 
system is allowed to evolve long enough to thermalize at $T(k)$. The temperature is 
then lowered according to a prescribed annealing schedule $T(k)$ until the energy of 
the system approaches it ground state value within an acceptable tolerance. It is 
important to keep the annealing rate slow enough in order to avoid trapping in some 
metastable states.

In our problem, the discrete configuration space includes the states of all possible 
pixel allocations. Starting from a given state (current configuration), a new state 
(new configuration) can be obtained by interchanging two arbitrarily selected pixels 
of different phases. This simple evolving procedure preserves the volume fraction of 
all involved phases and guarantees ergodicity in the sense that each state is 
accessible from any other state by a finite number of interchange steps. However, in 
the later stage of the procedure, biased and more sophisticated interchange rules, 
i.e., surface optimization, could be used to improve the efficiency. We choose the 
Metropolis algorithm as the acceptance criterion: the acceptance probability $P$ for 
the pixel interchange is given by 

\begin{equation}
\label{eq209}
P(E_{old}\rightarrow E_{new}) = \left\{{\begin{array}{*{20}c}1, \quad\quad \Delta E < 0, \\\\ \exp(-\Delta E/T),\quad \Delta E\ge 0,\end{array}}\right.
\end{equation}

\noindent where $\Delta E = E_{new}-E_{old}$. The temperature $T$ is initially chosen 
so that the initial acceptance probability for a pixel interchange with 
$\Delta E\ge 0$ averages approximately $0.5$. An inverse logarithmic annealing 
schedule which decreases the temperature according to $T(k) \sim 1/\ln(k)$ would in 
principle evolve the system to its ground state. However, such a slow annealing 
schedule is difficult to achieve in practice. Thus, we will adopt the more popular 
and faster annealing schedule $T(k)/T(0) = \lambda^k$, where constant $\lambda$ must 
be less than but close to one. This may yield suboptimal results, but, for practical 
purposes, will be sufficient. The convergence to an optimum is no longer guaranteed, 
and the system is likely to freeze in one of the local minima if the thermalization 
and annealing rate are not adequately chosen.

The two-point correlation function of a statistically homogeneous and isotropic 
medium is the focus of this paper. In this case, Eq.~(\ref{eq208}) reduces to

\begin{equation} 
\label{eq210}
E = \sum\limits_i \left[{S_2(r_i) - \widehat{S}_2(r_i)}\right]^2.
\end{equation}

\noindent Since for every configuration (structure), the corresponding two-point 
correlation function needs to be computed, the efficiency of the construction or 
reconstruction is mainly determined by the efficiency of the $S_2$-sampling algorithm.
 Furthermore, the properties of generated configurations (structures), i.e., isotropy 
of the medium, is also affected by the $S_2$-sampling algorithm. One of the most 
commonly used and efficient $S_2$-sampling algorithms is the 
\textit{orthogonal-sampling} algorithm introduced by Yeong and Torquato 
\cite{ 2PhysRevE.57.495, 3PhysRevE.58.224}. Due to the isotropic nature of the medium,
 every sampling direction should be equivalent. For simplicity, two orthogonal 
directions (usually the horizontal and vertical directions of a square lattice) are 
chosen and the two-point correlation function is sampled along these directions and 
averaged. At each pixel interchange, only the values of $S_2(r)$ sampled along the 
rows and columns that contain the interchange pixels are changed. Thus, the 
complexity of the algorithm is $O(N)$, where $N$ is the linear size of the system. 
However, for certain media with long-range correlations, the generated media have 
microstructures with two orthogonal anisotropic directions due to the biased sampling.
 Modifications of the orthogonal-sampling algorithm to preserve the isotropy of the 
underlying medium have been proposed, such as adding more sampling directions and 
using more isotropic lattices \cite{4cule:3428, 5sheehan:53}. Cule and Torquato i
ntroduced a new isotropy-preserving \textit{fast Fourier transform} (FFT) algorithm 
\cite{4cule:3428}. At each pixel interchange step, the two-point correlation function 
$S_2({\bf r})$ containing angle information is calculated in momentum space using an 
efficient FFT algorithm. Since information of all directions is considered, the 
generated media always have the required isotropy structures. However, since the 
complexity of FFT is $O(N\log_2 N)$, the algorithm is relatively time consuming.

We have developed an efficient and isotropy-preserving algorithm, namely, the 
\textit{Lattice-Point} algorithm by considering the black pixels as hard ``particles''
 on a particular lattice. The two-point correlation function is then computed in a 
similar way of obtaining the \textit{pair correlation function} $g_2(r)$ for an 
isotropic point process \cite{1torquato}. At each Monte-Carlo step, the randomly 
selected ``particle'' (black pixel) is given a random displacement subjected to the 
nonoverlapping constraint and the distances between the moved ``particle'' and all 
the other ``particles'' need to be recomputed. Thus the complexity of the algorithm 
is $O(N)$. Since all directions are effectively sampled, constructions based on the 
angle-averaged $S_2(r)$ well preserve isotropy of the media. A detailed discussion 
and applications of the algorithm will be included in the second paper of this series 
\cite{YJIAO}. In this paper, we only provide several illustrative examples generated 
from this algorithm. 

\section{Illustrative Examples}

As illustrative examples, we use the aforementioned (re)construction techniques to 
investigate both deterministic, crystal-like structures, and random systems. We also 
study a hypothetical medium with the two-point correlation function obtained from 
convex combination of known ones. In the case of completely deterministic structures, 
the algorithm produces almost perfect reconstructions. However, the optimization of 
disordered structures is significantly harder. Furthermore, we will see that for the 
media with long-range correlations, e.g., a damped-oscillating $S_2(r)$, the 
orthogonal-sampling algorithm may produce unexpected anisotropy, while the 
Lattice-Point algorithm well preserves isotropy of the media.

\subsection{Regular Array of Nonoverlapping Disks}

\begin{figure}[bthp]
\begin{center}
$\begin{array}{c}\\
\includegraphics[width=5cm,keepaspectratio]{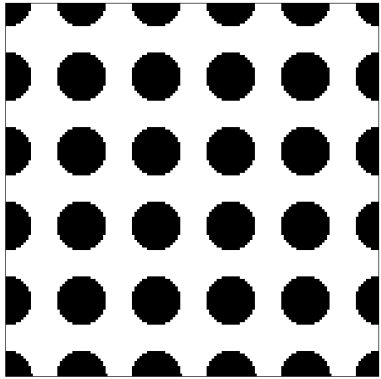}
\end{array}$
\end{center}
\caption{A realization of square array of nonoverlapping disks. The linear size of the system $N = 200$ 
pixels, volume fraction of black pixels $\phi_1 = 0.326$.}
\label{fig6}
\end{figure}

First, we consider specific two-dimensional and two-phase structure composed of a 
square array of nonoverlapping disks, as shown in Fig.~\ref{fig6}. This morphology 
may be viewed as a cross section of two-phase materials containing rod- or fiber-like 
inclusions. Various transport properties of these materials have been well explored 
because of their practical and theoretical importance in materials science 
\cite{15fiber}.

The regular structure is discretized by introducing an $N\times N$ square lattice. 
The volume fractions of black and white phases are $\phi_1 = 0.326$, $\phi_2 = 0.674$,
 respectively. The target two-point correlation of the digitized medium is sampled 
using both the orthogonal and the Lattice-Point algorithm for comparison purpose. The 
simulations start from random initial configurations (i.e., random collections of 
black and white pixels), at some initial temperature $T_0$, with fixed volume 
fractions $\phi_i$. At each Monte-Carlo (MC) step, when an attempt to exchange two 
randomly chosen pixels with different colors (or to randomly displace a chosen black 
pixel) is made, $S_2(r)$ is efficiently recomputed by using the orthogonal-sampling 
algorithm (or the Lattice-Point algorithm). The set of constants $\{\lambda_{MC}, 
\lambda_{tot}, \lambda\}$ specifies the annealing schedule: At each temperature, the 
system is thermalized until either $\lambda_{MC} N^2$ MC moves are accepted or the 
total number of attempts to change the original configurations reaches the value 
$\lambda_{tot} N^2$. Subsequently, the system temperature is decreased by the 
reduction factor $\lambda$, i.e., $T_{new} = \lambda T_{old}$.

\begin{figure}[bthp]
\begin{center}
$\begin{array}{c@{\hspace{1cm}}c}\\
\includegraphics[width=4cm,keepaspectratio]{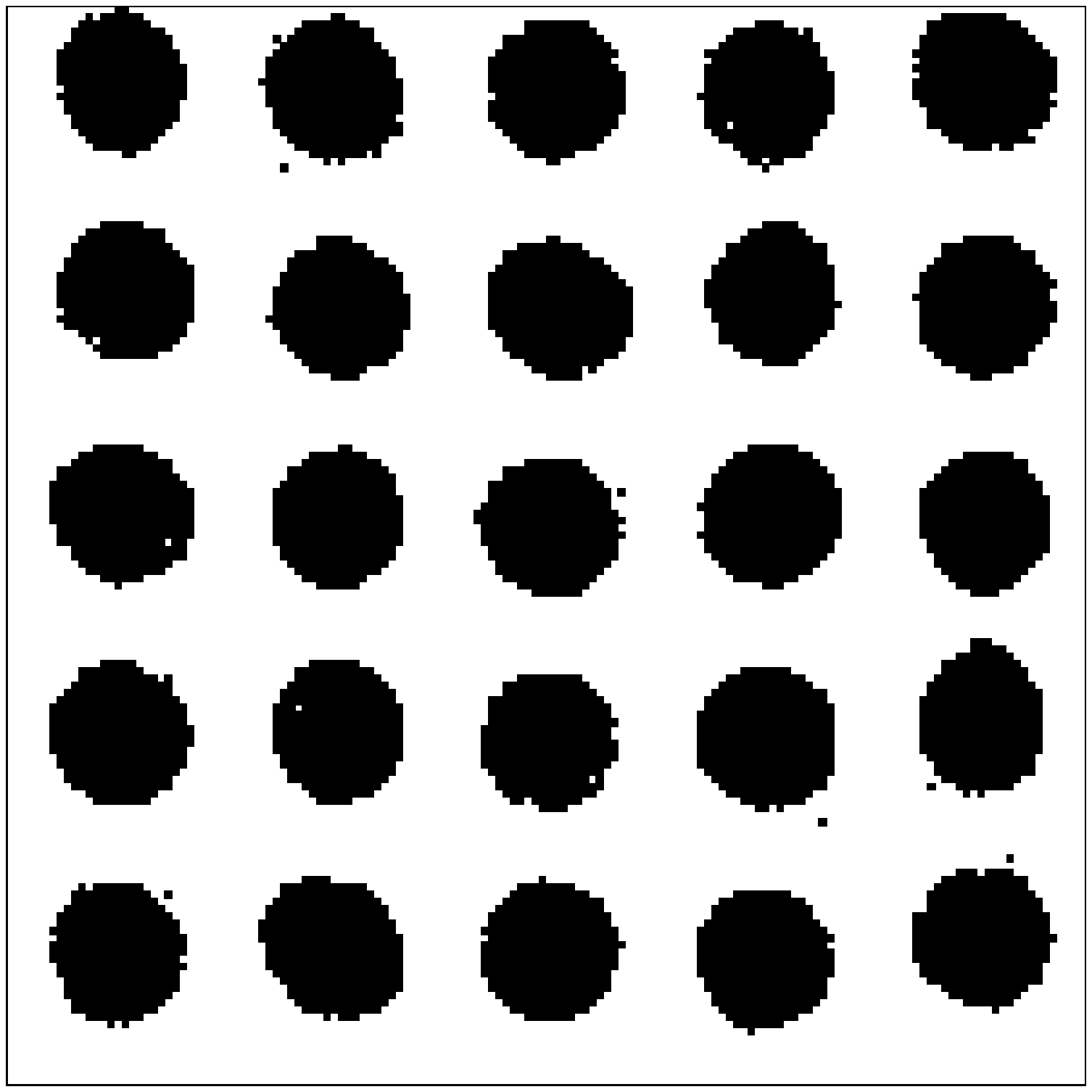} &
\includegraphics[width=4cm,keepaspectratio]{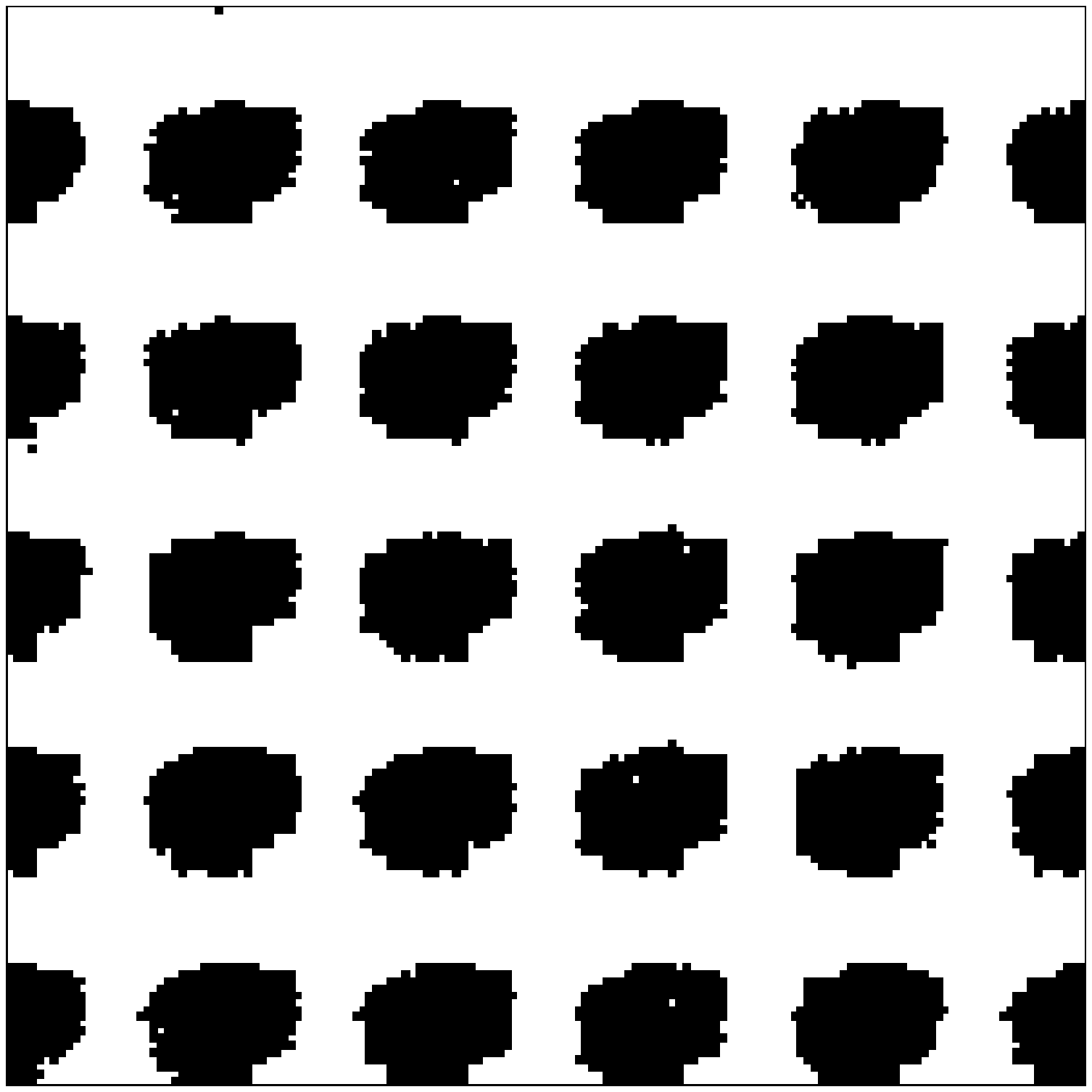} \\
\mbox{\bf (a)} & \mbox{\bf (b)}
\end{array}$
\end{center}
\caption{Reconstructed structures: (a) Square array of almost circular particles 
generated by the Lattice-Point algorithm. (b) Square array of particles generated 
by the orthogonal-sampling algorithm.}
\label{fig7}
\end{figure}

The reconstruction results are shown in Fig.~\ref{fig7}. Both of the algorithms are 
able to reproduce the exact global square-array arrangement of clusters of black 
pixels. This implies that the two-point correlation function of regular configurations
 contains enough structural information to properly characterize the long-range 
correlations. However, it is clear that the structure generated by the Lattice-Point 
algorithm has a better local arrangement of the pixels (i.e., the shape of the 
particles) than that generated by the orthogonal-sampling algorithm. This is because 
the orthogonal algorithm only uses structural information along two directions, 
which is not sufficient to reproduce detailed local structures, while the 
Lattice-Point algorithm efficiently uses information along all possible directions.

\subsection{Hypothetical Random Media with Long-Range Correlations}

In this example, we will generate two-dimensional statistically homogeneous and 
isotropic random media with long-range correlations (i.e., nontrivial inter-particle 
interactions). Examples of this type of media include low-density fluids and amorphous
 materials (i.e., porous media, randomly polymerized plastics, glass, etc.). A 
meaningful, yet nontrivial, two-point correlation function capturing these features is
 \cite{4cule:3428, 5sheehan:53, 7torquato:06}

\begin{equation}
\label{eq211}
\widehat{S}_2(r) = \phi_1^2 + \phi_1\phi_2 e^{-r/r_0}\frac{\sin(kr)}{kr},
\end{equation} 

\noindent where $k = 2\pi/a_0$. Here $r_0$ and $a_0$ are two characteristic length 
scales. The overall exponential damping is controlled by the correlation length $r_0$,
 determining the maximum correlations in the system. The constant $a_0$ determines 
oscillations in the term $\sin(kr)/(kr)$ which also decays with increasing $r$, such 
that $a_0$ can reduce the effective range of $r_0$. Interestingly, this hypothetical 
function is \textit{not} exactly realizable, because it violates the convexity 
condition Eq.~(\ref{eq117}) at the origin, or more generally the triangular inequality
 Eq.~(\ref{eq115}). However, we mainly focus on its damped-oscillating property and we
 will see that the algorithms are robust enough to detect violation of the convexity 
condition.

\begin{figure}[bthp]
\begin{center}
$\begin{array}{c@{\hspace{1cm}}c}\\
\includegraphics[width=4cm,keepaspectratio]{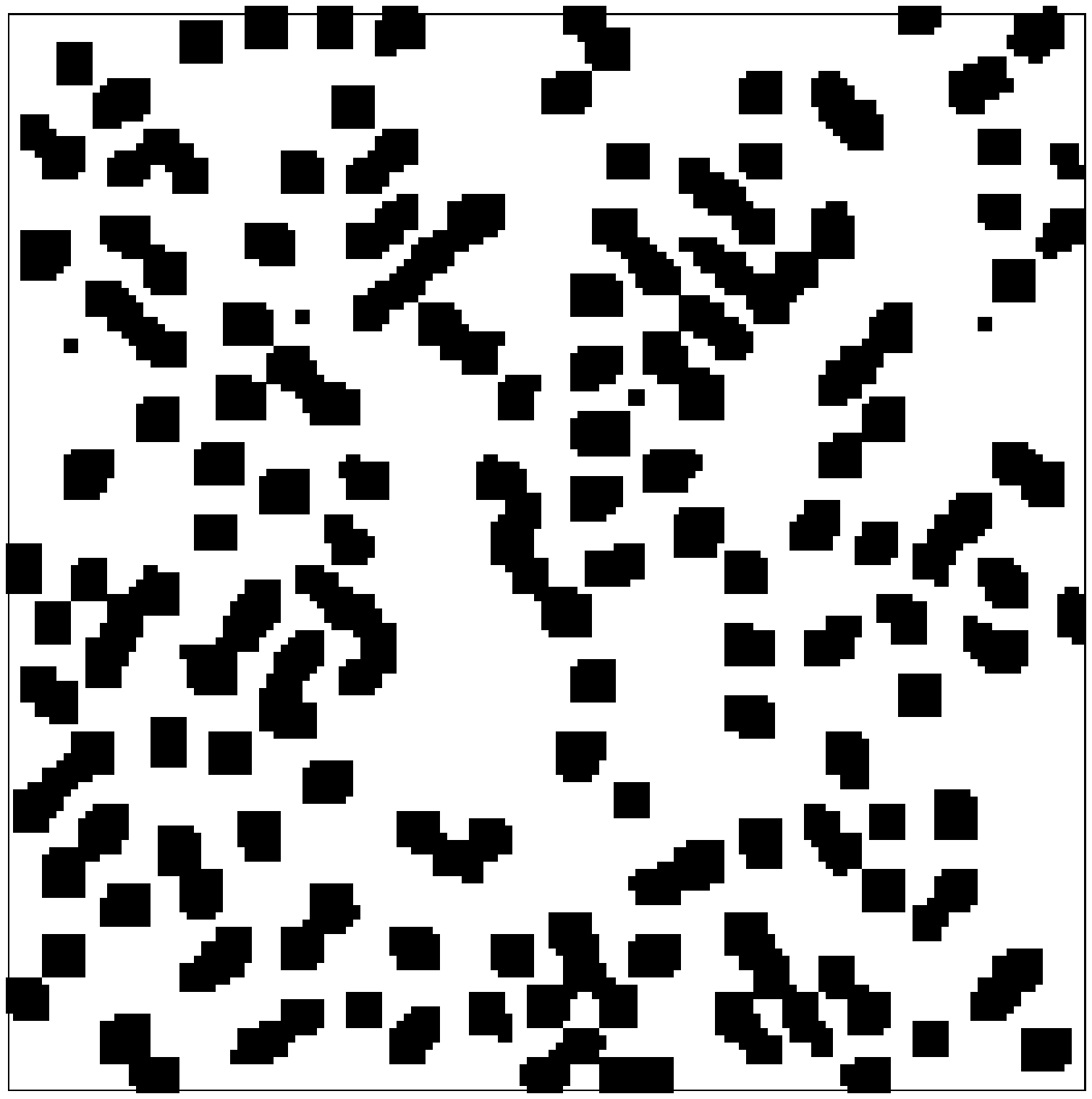} &
\includegraphics[width=4cm,keepaspectratio]{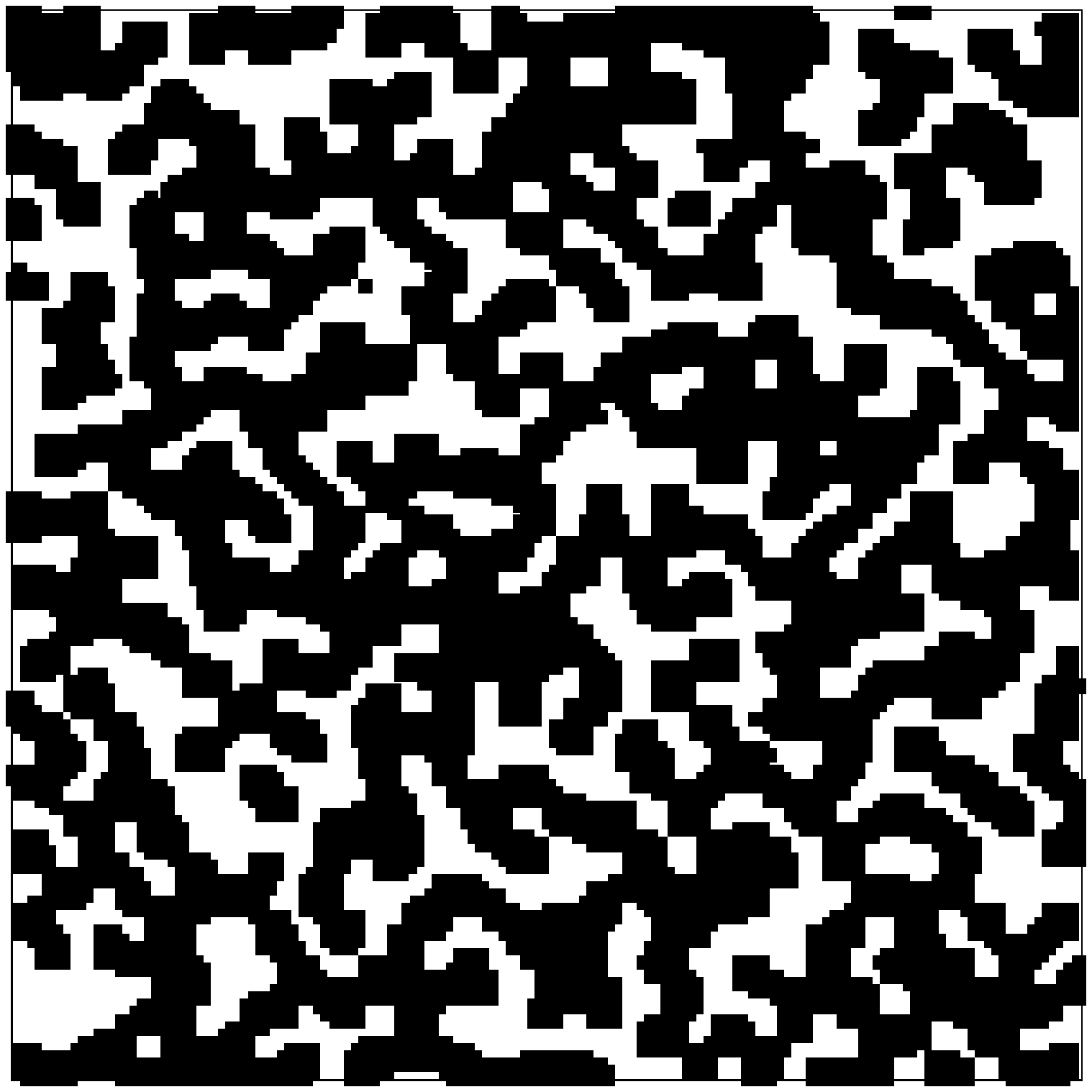} \\
\mbox{\bf (a1)} & \mbox{\bf (a2)}\\\\\\
\includegraphics[width=4cm,keepaspectratio]{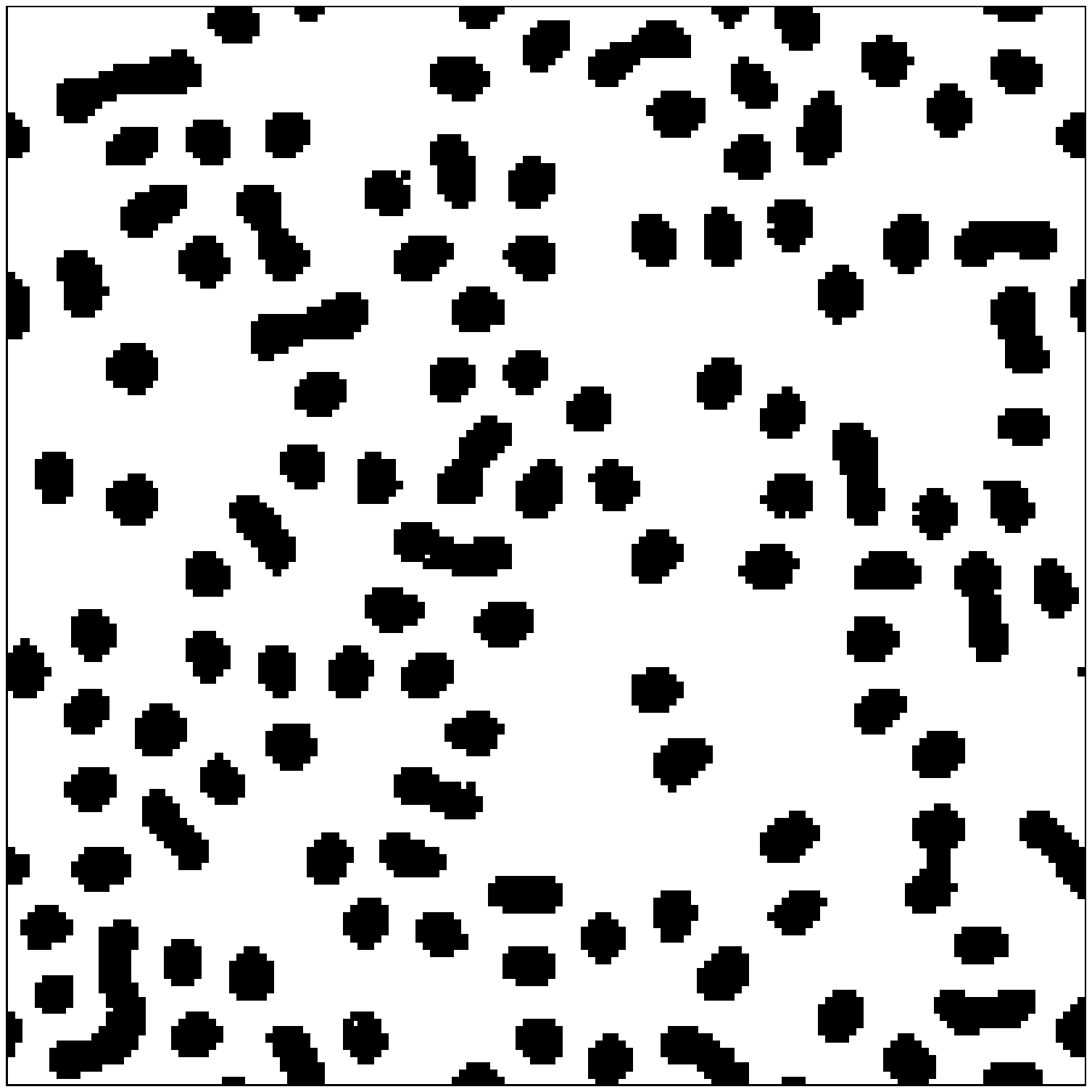} &
\includegraphics[width=4cm,keepaspectratio]{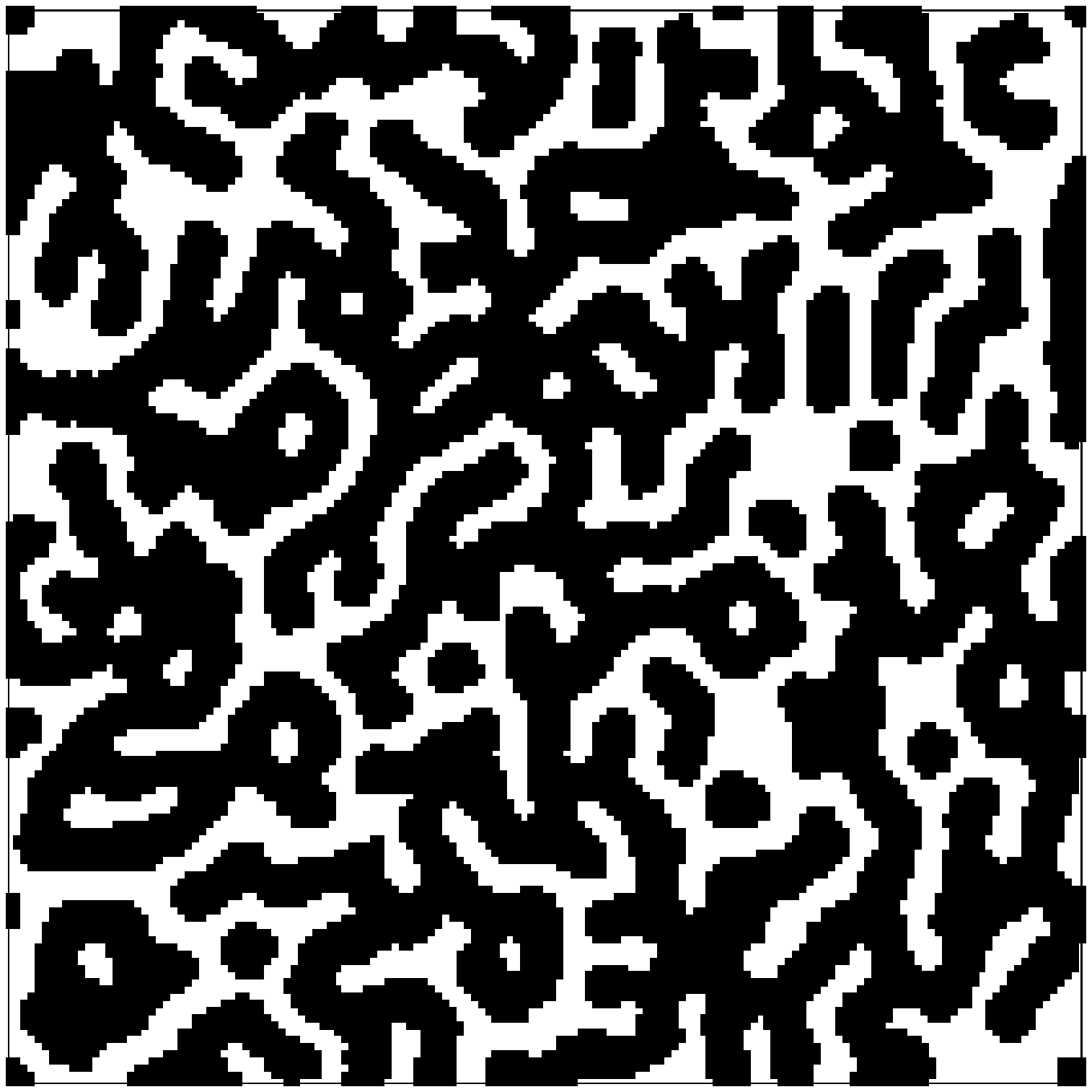} \\
\mbox{\bf (b1)} & \mbox{\bf (b2)}
\end{array}$
\end{center}
\caption{(a) Media with $\widehat{S}_2(r)$ given by Eq.~(\ref{eq211}) generated by 
the orthogonal-sampling algorithm: Left panel, volume fraction of black pixels 
$\phi_1 = 0.2$. Right panel, volume fraction of black pixels $\phi_1 = 0.5$. The 
linear size of the systems $N = 200$. (b) Media with $\widehat{S}_2(r)$ given by 
Eq.~(\ref{eq211}) generated by the Lattice-Point algorithm: Left panel, volume 
fraction of black pixels $\phi_1 = 0.2$. Right panel, volume fraction of black pixels 
$\phi_1 = 0.5$. The linear size of the systems $N = 200$.}
\label{fig8}
\end{figure}

For comparison purposes, both the orthogonal-sampling algorithm and the Lattice-Point 
algorithm are used in the construction, the results are shown in Fig.~\ref{fig8}. At a
 lower density of the black phase $\phi_1$, $a_0$ is manifested as a characteristic 
repulsion among different elements with diameter of order $a_0$. The repulsion 
vanishes beyond the length scale $r_0$. At a higher density, both length scales $a_0$ 
and $r_0$ are clearly noticeable in the distribution of the black and white phases. 
Note that the structures generated by the orthogonal-sampling algorithm exhibit some 
anisotropy features, i.e., containing stripes along $ \pm 45$ degree directions, which
 implies that the orthogonal-sampling algorithm should be used with care in the case 
where the medium has long-range correlations.

\begin{figure}[bthp]
\begin{center}
$\begin{array}{c}\\
\includegraphics[height=6cm,keepaspectratio]{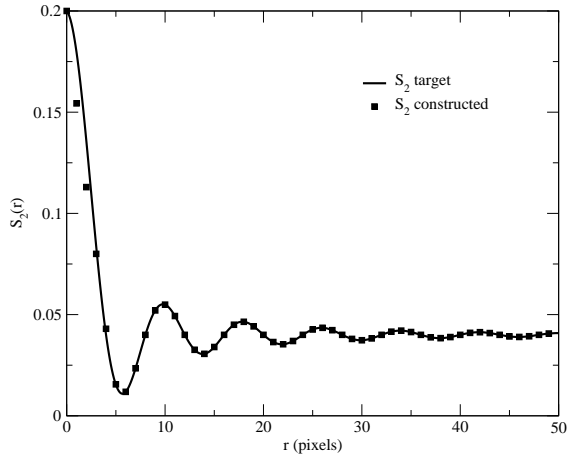}
\end{array}$
\end{center}
\caption{Target two-point correlation function given by Eq.~(\ref{eq211}) and that of 
constructed media with volume fraction $\phi_1 = 0.2$.}
\label{fig9}
\end{figure}

The target two-point correlation function $\widehat{S}_2(r)$ for $\phi_1 = 0.2$ and 
$S_2(r)$ sampled from generated structures are shown in Fig.~\ref{fig9}. It can be 
seen clearly that $\widehat{S}_2(r)$ is non-convex at the origin and the largest 
discrepancies between $S_2(r)$ and $\widehat{S}_2$ occur around the origin because 
$S_2(r)$ satisfies Eq.~(\ref{eq117}). This implies that our algorithms are robust 
enough and can be used to test realizability of hypothetical functions. Note that the 
following classes of functions \cite{7torquato:06}

\begin{equation}
f(r) = \exp\left[{-\left({\frac{r}{a}}\right)^\alpha}\right] \quad\quad \alpha > 1,
\end{equation}

\noindent and

\begin{equation}
f(r) = \frac{1}{\left[{1+(r/a)^2}\right]^{\beta -1}} \quad\quad  \beta \ge d,
\end{equation}

\noindent cannot correpond to a two-phase medium in $d$ dimensions also because of 
violation of triangular inequality Eq.~(\ref{eq121}).

\subsection{Hypothetical Random Media with Realizable Correlation Functions}

In this last example, we study a hypothetical statistically homogeneous and isotropic 
medium whose scaled autocovariance function is a convex combination of Debye random 
medium function $f_D(r)$ and damped-oscillating function $f_O(r)$, i.e.,

\begin{equation}
\label{eq212}
f(r) = \alpha_1 f_D(r) + \alpha_2 f_O(r), 
\end{equation}

\noindent where $\alpha_1 + \alpha_2 = 1$. Since both $f_D(r)$ and $f_O(r)$ are 
independent of volume fractions, the medium with scaled autocovariance function $f(r)$
 has \textit{phase-inversion symmetry} \cite{1torquato}, i.e., the structures with 
volume fraction of black pixels $\phi_1 = 0.2$ is statistically the same with those 
having volume fraction of white pixels $\phi_2 = 0.2$, if the colors of the two 
phases are inverted in the latter.

\begin{figure}[bthp]
\begin{center}
$\begin{array}{c}\\
\includegraphics[height=6cm,keepaspectratio]{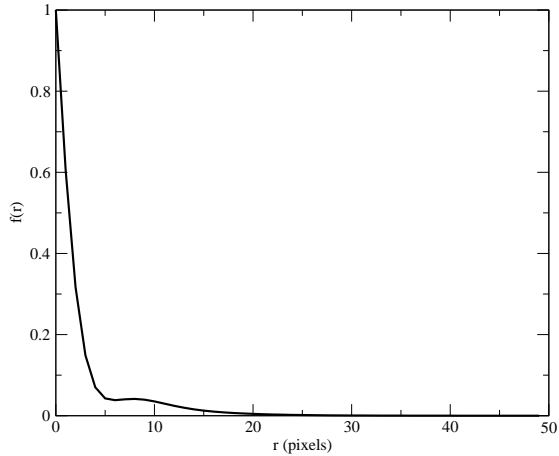}
\end{array}$
\end{center}
\caption{Combined scaled autocovariance function $f(r)$ with coefficients 
$\alpha_1 = 0.25, ~\alpha_2 = 0.75$.}
\label{fig10}
\end{figure}

\begin{figure}[bthp]
\begin{center}
$\begin{array}{c@{\hspace{1cm}}c@{\hspace{1cm}}c}\\
\includegraphics[height=3cm,keepaspectratio]{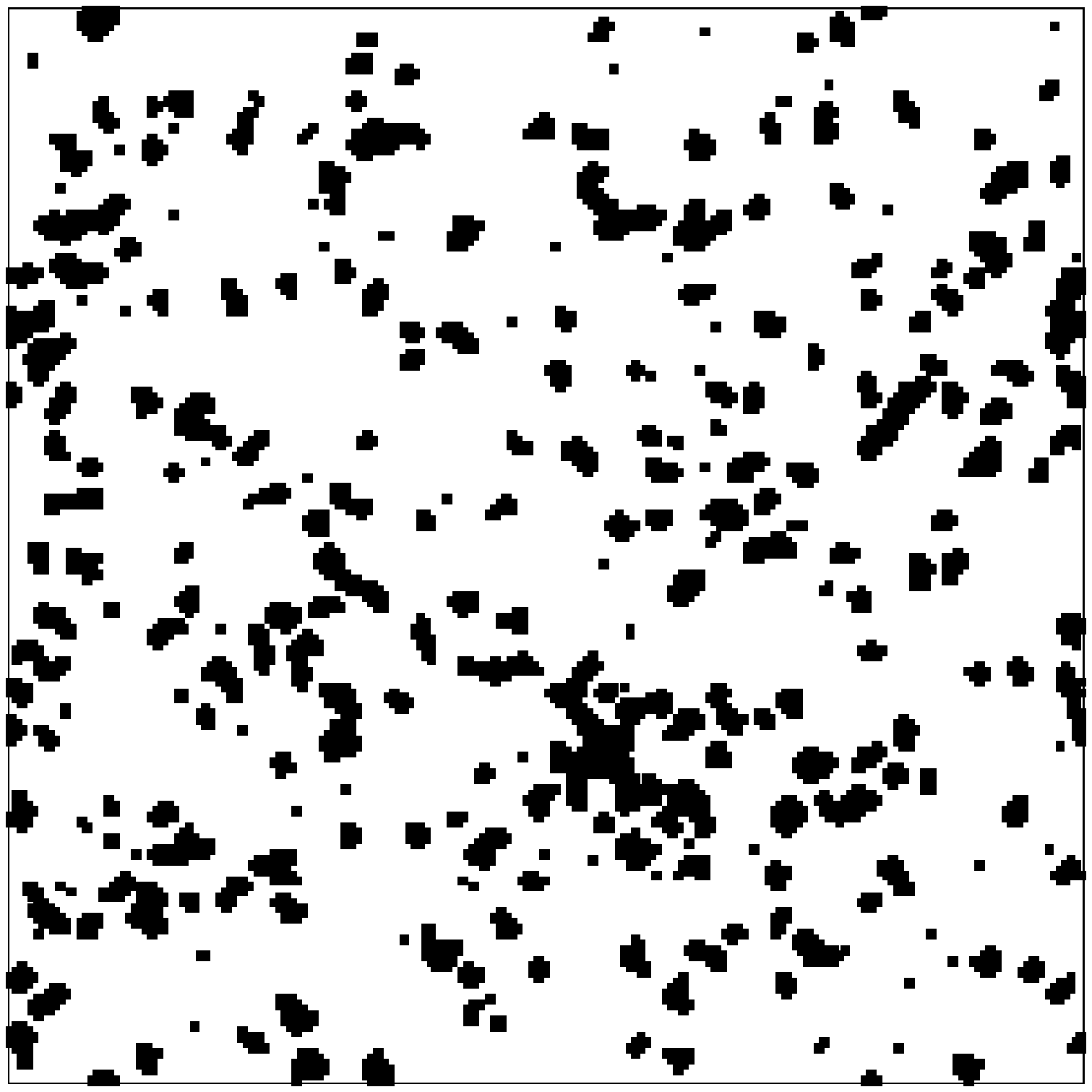} &
\includegraphics[height=3cm,keepaspectratio]{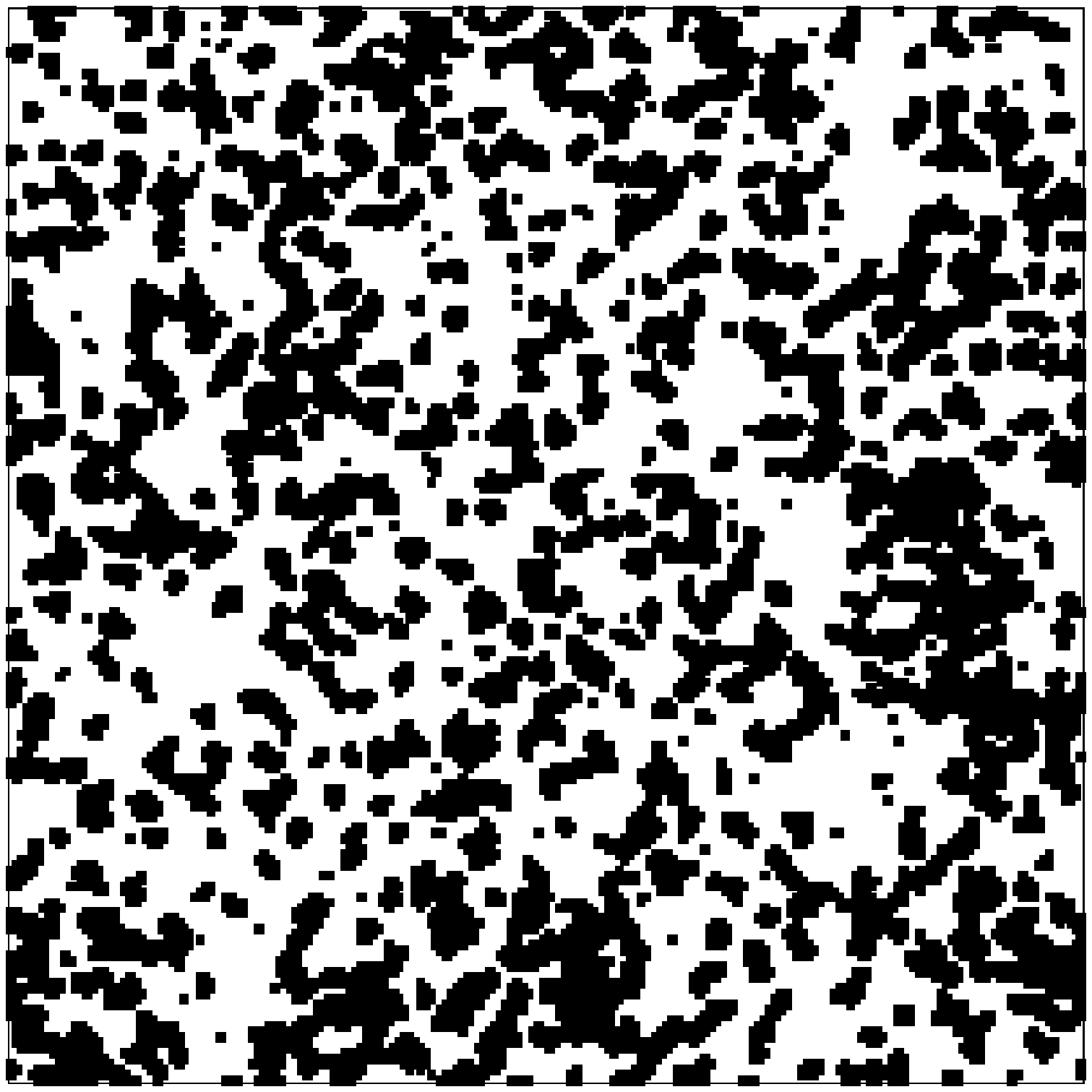} &
\includegraphics[height=3cm,keepaspectratio]{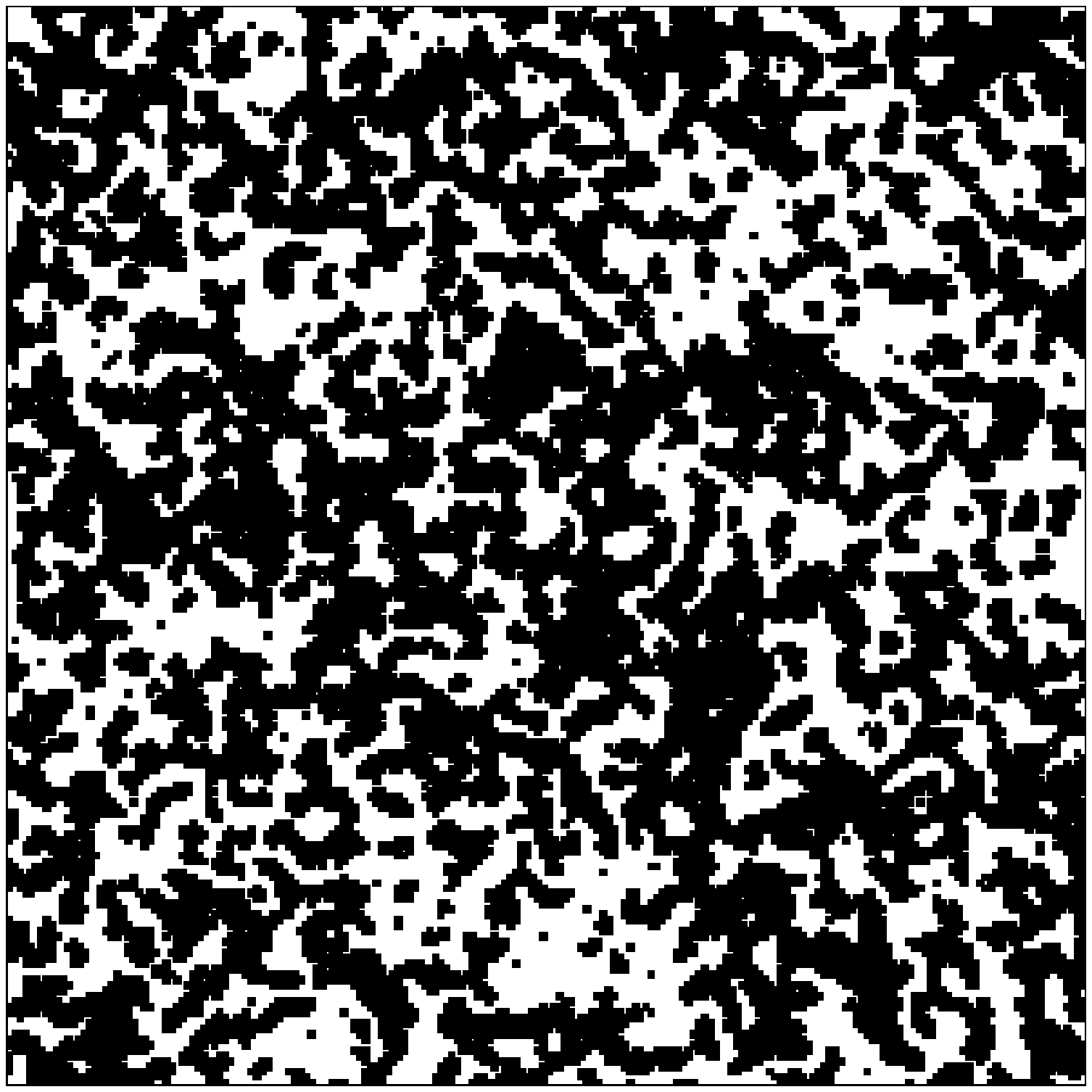} \\
\mbox{\bf (a)} & \mbox{\bf (b)} & \mbox{\bf (c)}
\end{array}$
\end{center}
\caption{Constructed media with scaled autocovariance function shown in 
Fig.~\ref{fig10}: (a) Volume fraction of black pixels $\phi_1 = 0.1$. (b) Volume 
fraction of black pixels $\phi_1 = 0.3$. (c) Volume fraction of black pixels 
$\phi_1 = 0.5$. The linear size of the systems $N = 200$.}
\label{fig11}
\end{figure}

\begin{figure}[bthp]
\begin{center}
$\begin{array}{c}\\
\includegraphics[height=6cm,keepaspectratio]{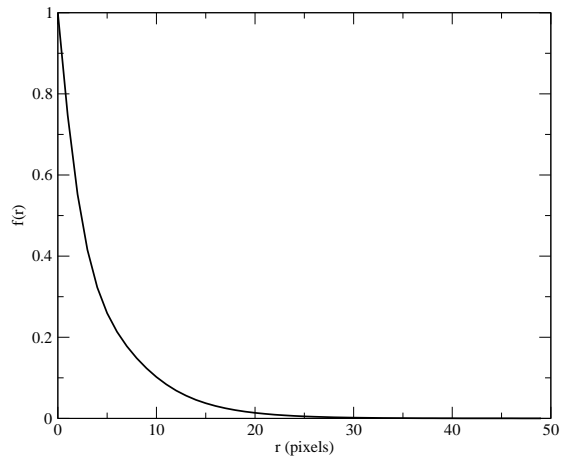}
\end{array}$
\end{center}
\caption{Combined scaled autocovariance function $f(r)$ with coefficients $\alpha_1 = 0.75, ~\alpha_2 = 0.25$.}
\label{fig12}
\end{figure}

\begin{figure}[bthp]
\begin{center}
$\begin{array}{c@{\hspace{1cm}}c@{\hspace{1cm}}c}\\
\includegraphics[height=3cm,keepaspectratio]{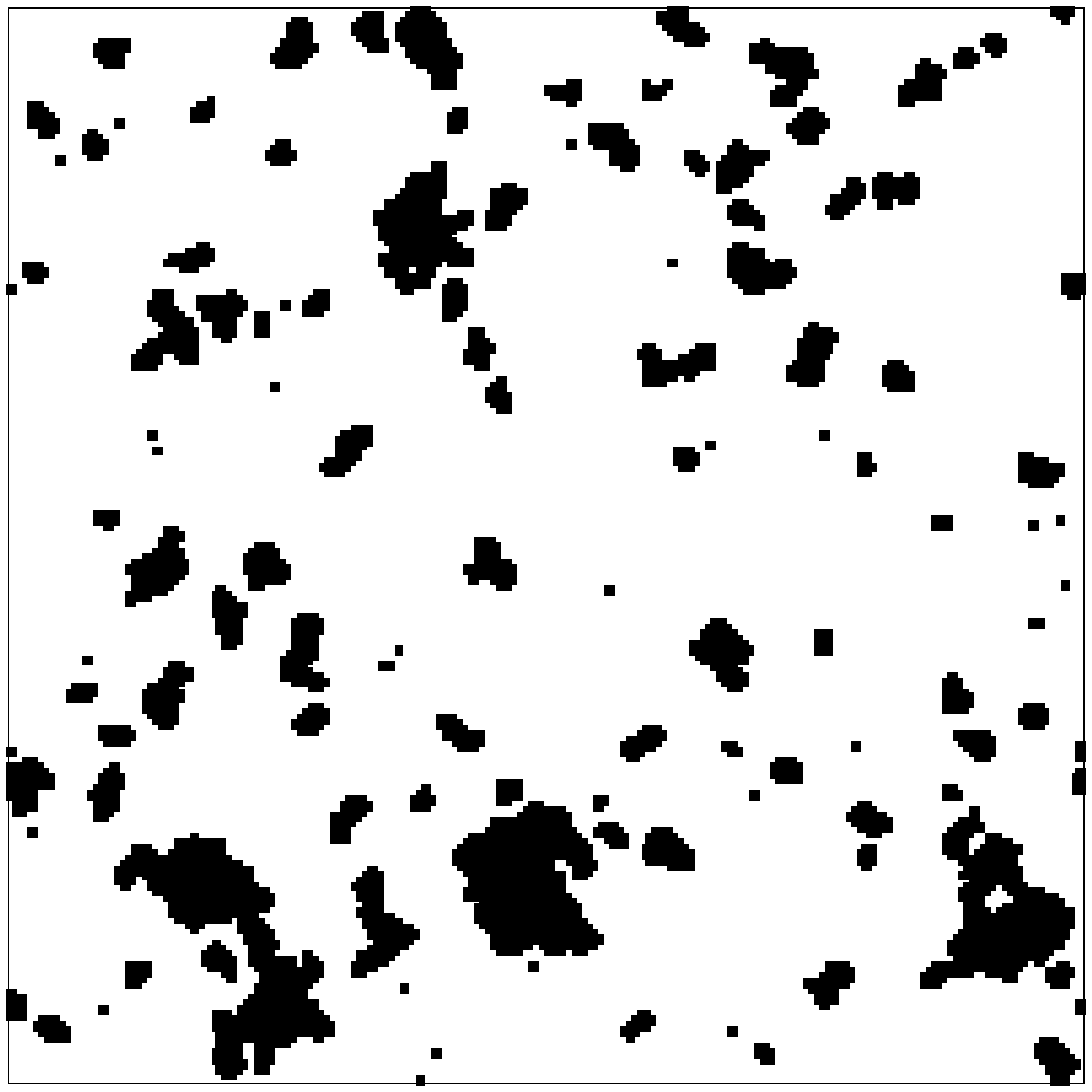} &
\includegraphics[height=3cm,keepaspectratio]{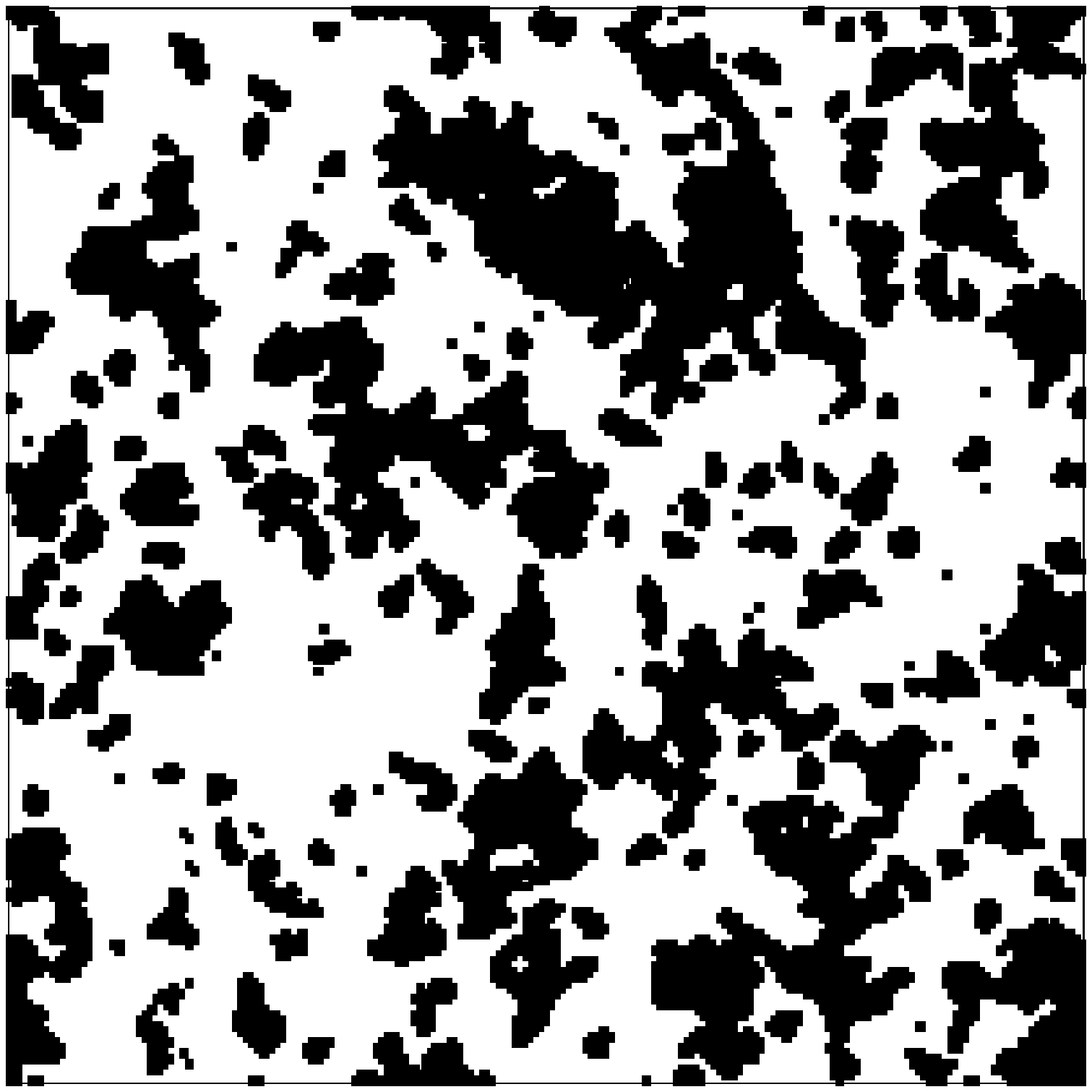} &
\includegraphics[height=3cm,keepaspectratio]{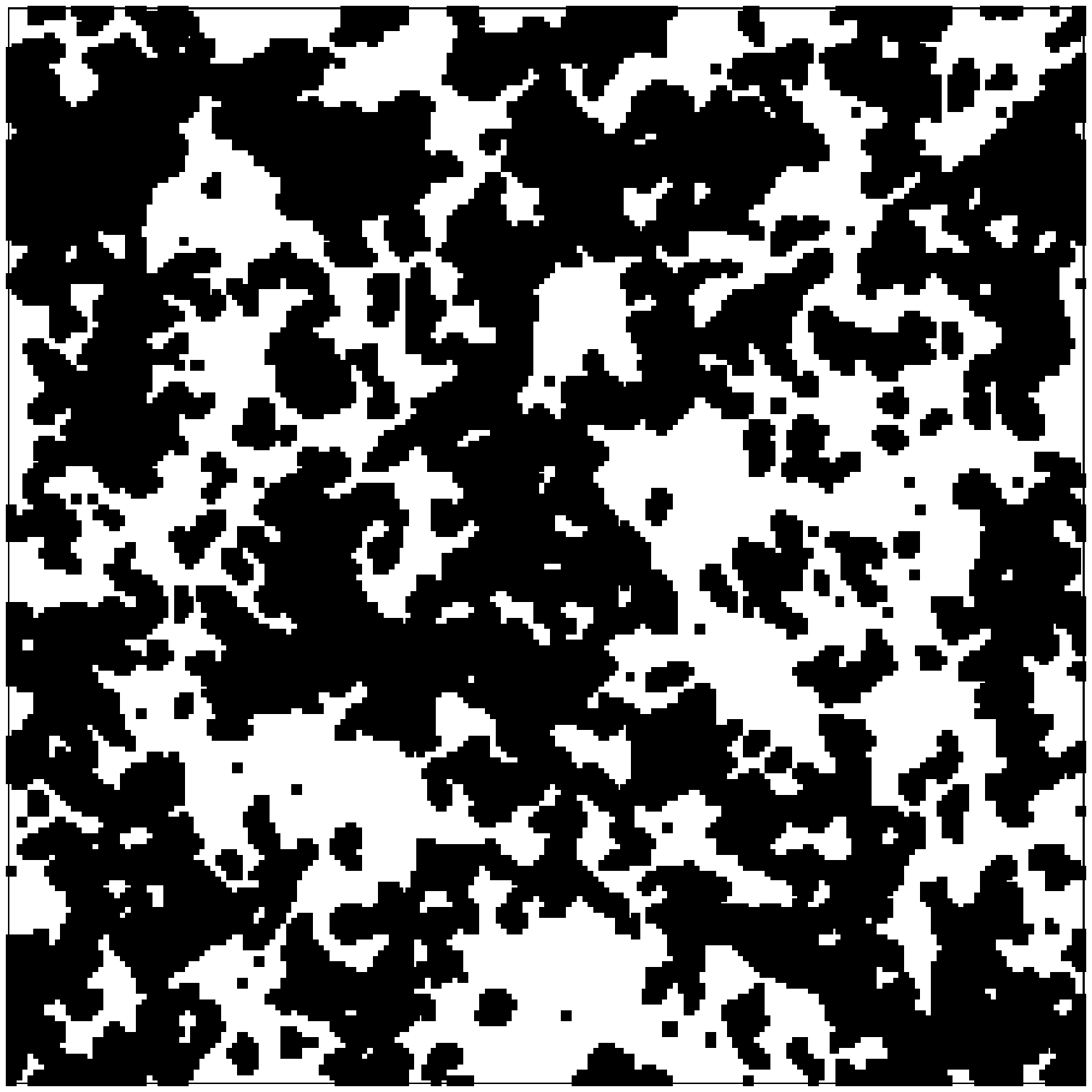} \\
\mbox{\bf (a)} & \mbox{\bf (b)} & \mbox{\bf (c)}
\end{array}$
\end{center}
\caption{Constructed media with scaled autocovariance function shown in Fig.
~\ref{fig12}: (a) Volume fraction of black pixels $\phi_1 = 0.1$. (b) Volume fraction 
of black pixels $\phi_1 = 0.3$. (c) Volume fraction of black pixels $\phi_1 = 0.5$. 
The linear size of the systems $N = 200$.}
\label{fig13}
\end{figure}

Different constant pairs $(\alpha_1,~ \alpha_2)$ can be used to construct $f(r)$ with 
required properties. In particular, we choose two pairs: $(0.25,~ 0.75)$ and 
$(0.75,~ 0.25)$. The construction results obtained by application of the Lattice-Point
 algorithm are shown in Figs.~\ref{fig10},~\ref{fig11} and Figs.~\ref{fig12},
~\ref{fig13}. For $\alpha_1 = 0.25, ~\alpha_2 = 0.75$, $f_O(r)$ is dominant in the 
combination. At lower densities, the generated structures resemble those with ``pure''
 damped oscillationg two-point functions, i.e., dispersions of particles, altough they
 contain more clusters. At higher densities, some stripe-like structures and several 
(almost equal-sized) clusters can be identified. For $\alpha_1 = 0.75, 
~\alpha_2 = 0.25$, $f_D(r)$ is dominant in the combined $f(r)$. Clusters with 
considerable sizes form even at low densities, which is a consequence of the large 
effective correlation length in $f_D(r)$. However, no stripe-like structures can be 
identified in the generated structures, since the contribution of $f_O(r)$ is 
significantly supressed by its damping nature and the small combination constants.

The results imply that even a simple combination of two basis functions enables one 
to obtain scaled autocovariance functions with properties of interest and to generate 
a variety of structures with controllable morphological features, e.g., local 
``particle'' shape and cluster size.  

\section{Conclusions}

In this paper, we have provided a general rigorous scheme to model and categorize 
two-phase statistically homogeneous and isotropic media. In particular, given a set of
 basis functions, we have shown that the medium can be modeled by a map $\wp$ composed
 of convex combination and product operations. The basis functions should be 
realizable but, if they are not, they should at least satisfy all the known necessary 
conditions for a realizable autocovariance function. We have gathered all the known 
necessary conditions and made a conjecture on a possible new condition based on 
simulation results. A systematic way of determining basis functions is not available 
yet. We proposed a set of basis functions with simple analytical forms that capture 
salient microstructural features of two-phase random media.

We give for the first time a rigorous mathematical formulation of the (re)construction
 problem and showed that the two-point correlation function alone cannot completely 
specify a two-phase heterogeneous material. Moreover, we devised an efficient and 
isotropy-preserving (re)construction algorithm, namely, the Lattice-Point algorithm 
to generate realizations of materials based on the Yeong-Torquato technique. We also 
provided an example of non-realizable yet non-trivial two-point correlation function 
and showed that our algorithm can be used to test realizability of hypothetical 
functions. An example of generating hypothetical random media with combined realizable
 correlation functions was given as an application of our general scheme. We showed 
that even a simple combination of two basis functions enables one to produce media 
with a variety of microstructures of interest and therefore a means of categorizing 
microstructures.

We are investigating applications of our general scheme in order to model real 
materials. We are also developing more efficient (re)construction algorithms. There is
 a need for a theoretical and numerical analysis of the \textit{energy threshold} of 
the algorithm, which is the aforementioned ``acceptable tolerance''. This quantity 
provides an indication of the extent to which the algorithms have reproduced the 
target structure and it is directly related to the non-uniqueness issue of 
reconstructions \cite{2PhysRevE.57.495, 3PhysRevE.58.224, 4cule:3428, 5sheehan:53}. 
Moreover, additional realizable basis functions are needed to construct a complete 
basis set. Such work will be reported in our future publications.

\appendix

\section{}

The complementary error function $f_{ce}(r)$ is defined as

\begin{equation}
\label{eqa1}
f_{ce}(r) = \frac{2}{\sqrt{\pi}}\int_{r/a}^{\infty}e^{-t^2}dt,
\end{equation}

\noindent where $r \ge 0$ and $a$ is the effective correlation length. It is easy to 
check that $f_{ce}(r)$ satisfies the known necessary conditions collected in this 
paper except for Eq.~(\ref{eq118}), which can only be checked for a finite number of 
cases. Realizations of the random medium associated with $f_{ce}(r)$ have been 
constructed using the Yeong-Torquato technique with very high numerical precision. 
Thus, we believe $f_{ce}(r)$ to be a valid candidate for a realizable scaled 
autocovariance function.


\begin{acknowledgments}
Acknowledgment is made to the Donors of the American Chemical Society Petroleum 
Research Fund for support of this research. 
\end{acknowledgments}

\end{document}